\documentclass[showpacs,twocolumn,preprintnumbers,amsmath,amssymb]{revtex4}
\usepackage{amsmath,amsfonts,latexsym,amssymb,graphicx,graphics,epsfig,subfigure,color,makeidx}
\usepackage{xcolor,diagbox}
\usepackage{multirow}
\usepackage[colorlinks,linkcolor=blue,anchorcolor=blue,citecolor=green,urlcolor=blue]{hyperref}
\usepackage{mathrsfs}
\usepackage{amsmath}
\newcommand {\nn}{\nonumber}

\begin{document}
\title{Quasinormal modes and echoes of a double braneworld}

	\author{Qin Tan$^{a}$}
\author{Sheng Long$^{a}$}
\author{Weike Deng$^{a}$}
\author{Jiliang Jing$^{a}$}\email{jljing@hunnu.edu.cn, corresponding author.}

\affiliation{$^{a}$Department of Physics, Key Laboratory of Low Dimensional Quantum Structures and Quantum Control of Ministry of Education, Synergetic Innovation Center for Quantum Effects and Applications, Hunan Normal University, Changsha, 410081, Hunan, China}

\begin{abstract}
In this work, we study the gravitational quasinormal modes and the gravitational echoes of a double braneworld. The double braneworld is a kind of split thick brane, which is crucial for addressing the hierarchy problem in the thick brane scenarios. Using the Bernstein spectral method, direct integration method, and asymptotic iteration method, we calculate the quasinormal mode frequencies of the double brane. We find that the quasinormal spectrum is very different from that of the single brane model, especially the high overtone mode. We also perform numerical evolution to study the time-domain properties of the characteristic modes of the double brane. The results show that when the degree of brane splitting is large, gravitational echoes of oscillation attenuation between sub-branes will appear in the thick brane. Furthermore, different long-lived Kaluza-Klein modes interfere with each other, resulting in a beating effect. Compared to a single brane model, the phenomenon of the split double brane is richer, and the lifetime of the massive Kaluza-Klein graviton of the double brane is longer. These phenomena may have potential phenomenological interest. We also explore the gravitational wave signature of the KK graviton of the thick brane and find that the violent splitting of the brane may cause the corresponding frequency to fall within the detection range of the high-frequency gravitational wave detector.

\end{abstract}

\maketitle

\section{Introduction}
\label{Introduction}

Braneworld models have been a topic of extensive research for many years, offering a new perspective on the nature of spacetime and providing a novel approach to address the hierarchy problem between the Planck and Electroweak scales~\cite{ArkaniHamed:1998rs,Antoniadis:1998ig,Randall:1999ee}. The Randall-Sundrum (RS) models, which involve branes embedded in a five-dimensional anti-de Sitter spacetime, have been particularly influential in this regard. The RS-I model~\cite{Randall:1999ee} proposes a solution to the hierarchy problem by introducing two branes, while the RS-II model~\cite{Randall:1999vf} extends this concept by pushing one brane to infinity. A remarkable feature of the RS-II model is the recovery of the four-dimensional Newtonian potential, even with an infinite extra dimension. These brane models have found wide-ranging applications in various areas of physics, including black hole physics, particle physics, and cosmology~\cite{Shiromizu:1999wj,Tanaka:2002rb,Gregory:2008rf,Jaman:2018ucm,Adhikari:2020xcg,Geng:2020fxl,Geng:2021iyq,Geng:2022dua,Bhattacharya:2021jrn,ValeixoBento:2022qca}.

While the RS-II thin brane model neglects the thicknesses and internal structure of the brane, treating its energy density as a delta function along the extra dimension, investigating the brane's inner structure requires considering thick brane models. These models, proposed by DeWolfe et al.\cite{DeWolfe:1999cp,Gremm:1999pj,Csaki:2000fc}, combine the domain wall model without gravity\cite{Akama:1982jy,Rubakov:1983bb} with the RS-II model to describe thick branes generated by one or more matter fields. Extensive research has been conducted on thick brane solutions in various gravity theories, as well as the localization of the gravitational zero mode and various matter fields on the brane~\cite{Afonso:2007gc,Dzhunushaliev:2010fqo,Dzhunushaliev:2011mm,Geng:2015kvs,Melfo2006,Almeida2009,Zhao2010,Chumbes2011,Liu2011,Bazeia:2013uva,Xie2017,Gu2017,ZhongYuan2017,ZhongYuan2017b,Zhou2018,Hendi:2020qkk,Xie:2021ayr,Moreira:2021uod,Xu:2022ori,Silva:2022pfd,Xu:2022gth}. For more comprehensive information on the braneworld see these reviews~\cite{Dzhunushaliev:2009va,Maartens:2010ar,Liu:2017gcn,Ahluwalia:2022ttu}. In addition to the zero mode, the existence of massive Kaluza-Klein (KK) particles on the brane, which are beyond the standard model. These massive KK modes may not stay on the brane forever, but escape into extra dimensions. Therefore, for these modes, the brane is dissipative. In dissipative systems, these characteristic modes is known as quasinormal modes (QNMs). It has garnered significant attention in various fields, especially in black hole physics~\cite{Berti:2009kk,Kokkotas:1999bd,Nollert:1999ji,Konoplya:2011qq,Cardoso:2016rao,Jusufi:2020odz,Cheung:2021bol}. Recent studies have shown that there may exist a set of discrete modes, known as QNMs of a brane, in both thin and thick braneworld scenarios~\cite{Seahra:2005wk,Seahra:2005iq,Tan:2022vfe,Tan:2023cra,Tan:2024url,Jia:2024pdk,Tan:2024aym}. Investigating these QNMs can provide a deeper understanding of the properties and dynamics of branes. Detecting these modes would open up new ways to understand the nature of spacetime.

One particularly interesting aspect of thick brane models is the phenomenon of brane splitting, where a single thick brane splits into two or more sub-branes~\cite{Melfo:2002wd,Castillo-Felisola:2004omi,deBrito:2014pqa,deSouzaDutra:2014ddw,Farokhtabar:2016fhm,Xie:2019jkq}. One fascinating property of the split brane is that it may be possible to solve hierarchical problem simultaneously in the framework of an RS-II model~\cite{Guerrero:2006gj,Ahmed:2012nh,deSouzaDutra:2013rwa}. This is the lack of the original RS-II model and the single thick brane models. This scenario offers a rich platform for exploring the properties of thick branes and their associated QNMs. Moreover, the presence of multiple sub-branes in a brane-splitting scenario may give rise to fascinating phenomena, such as gravitational echo. Gravitational echo, which are delayed gravitational wave signals, have been proposed as a potential signature of exotic compact objects and modified gravity theories~\cite{Cardoso:2017cqb,Jaramillo:2020tuu,Witek:2012tr,Cardoso:2019rvt,Mark:2017dnq,Conklin:2017lwb,Barcelo:2017lnx,Qian:2024zvq,Lin:2023qgd}. In the context of thick branes, gravitational echoes may emerge due to the interaction of gravitational perturbations with the multiple sub-branes in a brane-splitting scenario~\cite{Zhu:2024gvl}.

In this paper, we aim to investigate the QNMs and potential gravitational echo in a thick brane model with internal structure, focusing specifically on brane-splitting scenarios. By studying the properties of the QNMs, we seek to gain insights into the structure and stability of the brane. Furthermore, we will explore the possibility of gravitational echoes in the brane-splitting scenario. Our findings may offer new perspectives on detecting extra dimensions and probing the nature of gravity in braneworld models.

The rest of this paper is organized as follows. In Sect.~\ref{BRANE WORLD MODEL}, we introduce the thick brane model with internal structure and brane-splitting, and derive the equations governing the gravitational perturbations. In Sect.~\ref{quasinormal modes and echo of the branes}, we use semi-analytic method to solve the quasinormal frequency of the split thick brane in the frequency domain, and further study the QNMs of the thick brane and the properties of the gravitational echo in the time domain by numerical evolution. Finally, the conclusions and discussions are given in Sect.~\ref{Conclusion}.

\section{Braneworld model in general relativity}
\label{BRANE WORLD MODEL}
In this section, we review the thick brane solution in five-dimensional general relativity briefly and derive the perturbed equation for the transverse-traceless gravitational perturbation. A thick brane can be generated by various matter fields like scalar fields and vector fields. We choose that the thick brane is generated by a canonical scalar field. Thus, the action of the thick brane is
\begin{eqnarray}
	S=\int d^5x\sqrt{-g}\left(\frac{1}{2\kappa^{2}_{5}}R-\frac{1}{2}g^{MN}\partial_{M}
	\varphi\partial_{N}\varphi -V(\varphi)\right),\label{action}
\end{eqnarray}
where $\kappa_{5}^{2}=8\pi G_{5}$ is the five-dimensional gravitational coupling constant, which is set to $\kappa_{5}=1$ in this paper. The Einstein equation and the equation of motion for the scalar field are
\begin{eqnarray}
	R_{MN}-\frac{1}{2}Rg_{MN}&=&-\frac{1}{2}g_{MN}\left(\partial^{A}\varphi\partial_{A}\varphi-V(\varphi)\right) \nonumber\\
	&&+\partial_{M}\varphi\partial_{N}\varphi,\label{field equation}\\
	g^{MN}\nabla_{M}\nabla_{N}\varphi&=&\frac{\partial V(\varphi)}{\partial\varphi}.\label{motion equation}
\end{eqnarray}
Throughout this paper, indices $M,N,\dots=0,1,2,3,5$ and $\mu,\nu\dots=0,1,2,3$ denote the bulk and the brane coordinates, whereas indices $i,j\dots=1,2,3$ label the three-dimensional space ones on the brane.

\begin{figure}
	\subfigure[~$\delta=1$]{\label{figw123P1}
		\includegraphics[width=0.22\textwidth]{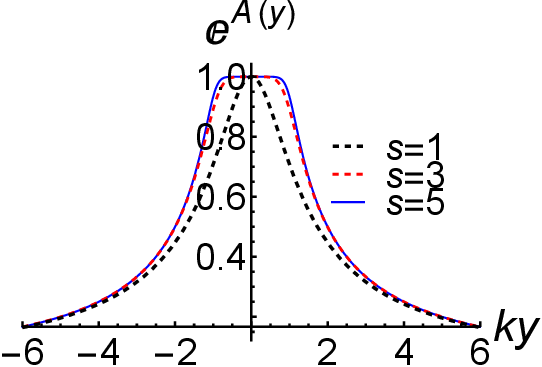}}
	\subfigure[~$s=3$]{\label{figw456P2}
		\includegraphics[width=0.22\textwidth]{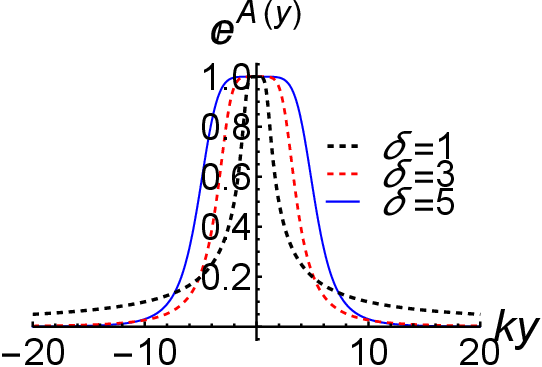}}
	\caption{The shapes of the warp factor~\eqref{warpfactorsolution1}.}\label{figwarpfactor}
\end{figure}

\begin{figure}
	\subfigure[~$\delta=1$]{\label{figphiP1}
		\includegraphics[width=0.22\textwidth]{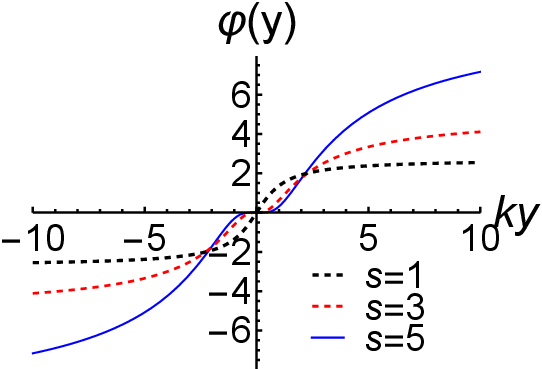}}
	\subfigure[~$s=3$]{\label{figphiP2}
		\includegraphics[width=0.22\textwidth]{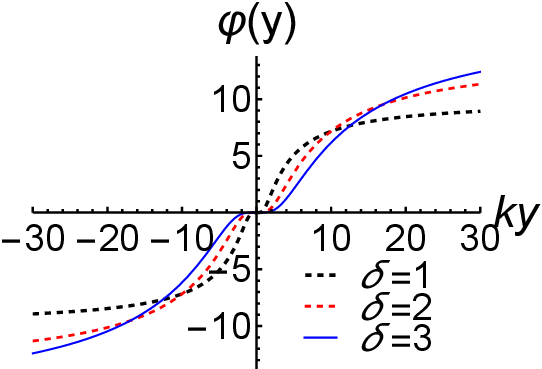}}
	\caption{Plots of the scalar field~\eqref{scalarfieldsolution1}.}\label{figphi}
\end{figure}

\begin{figure}[htbp]
	\subfigure[~$\delta=1$]{\label{figVphiP1}
		\includegraphics[width=0.22\textwidth]{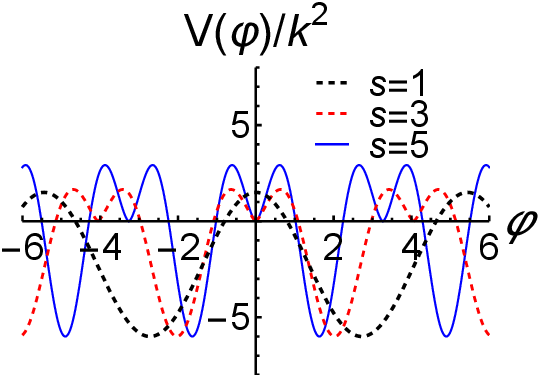}}
			\subfigure[~$s=3$]{\label{figVphiP2}
			\includegraphics[width=0.22\textwidth]{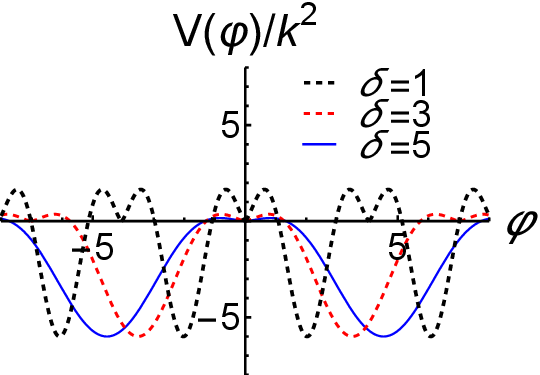}}
	\caption{Plots of the scalar potential~\eqref{scalarpotentialsolution1}.}\label{figVphi}
\end{figure}

\begin{figure}
	\subfigure[~$\delta=1$]{\label{figrhoP1}
		\includegraphics[width=0.22\textwidth]{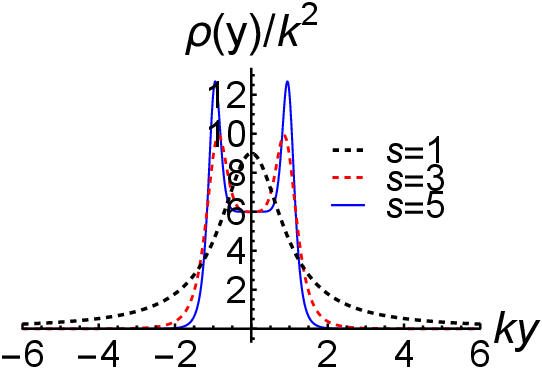}}
	\subfigure[~$s=3$]{\label{figrhoP2}
		\includegraphics[width=0.22\textwidth]{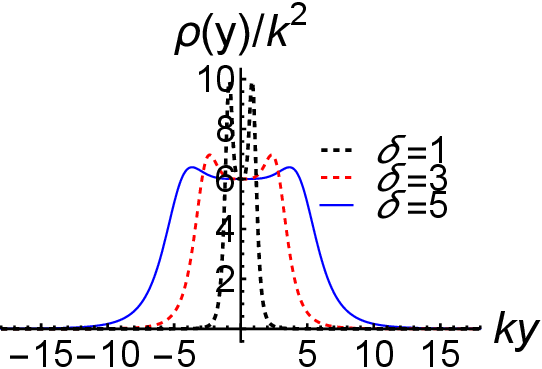}}
	\caption{Plots of the energy density~\eqref{energydenisty}.}\label{figrho1}
\end{figure}

By transform to the so-called “gauge” coordinates, a static flat brane metric can be written as~\cite{Melfo:2002wd}
\begin{equation}
	ds^2=e^{2A(y)}\eta_{\mu\nu}dx^\mu dx^\nu+e^{2H(y)}dy^2,
	\label{metric}
\end{equation}
where the four-dimensional Minkowski metric $\eta_{\mu\nu}$ is $\eta_{\mu\nu}=\text{diag}(-1,1,1,1)$. Combining the metric~\eqref{metric}, Einstein equation~\eqref{field equation}, and the equation of motion for the scalar field~\eqref{motion equation}, the specific dynamical equations are
\begin{eqnarray}
	2A'^2-A'H' +A''&=&-\frac{1}{2}\varphi'^2-V,  \label{EoMs1}\\
	6A'^2&=&\frac{1}{2}\varphi'^2-V,  \label{EoMs2}\\
	\varphi{''}+4A'\varphi'&=&\frac{\partial V}{\partial\varphi},  \label{EoMsphi}
\end{eqnarray}
where prime is the derivative with respect to $y$. The thick brane solution was investigated in Ref.~\cite{Melfo:2002wd}:
\begin{eqnarray}
	A(y)&=&\delta H(y)=\frac{-\delta}{2s}\ln\left[1+\left(\frac{ky}{\delta}\right)^{2s}\right] \label{warpfactorsolution1},\\
	\varphi(y)&=&\frac{\sqrt{3\delta(2s-1)}}{s}\arctan\left(\frac{ky}{\delta}\right),\label{scalarfieldsolution1}\\
	V(\varphi)&=&3k^{2}\left[\frac{2s+4\delta-1}{2\delta}\cos\left(\frac{s\varphi}{\sqrt{3\delta(2s-1)}}\right)^{2}-2\right]\nonumber\\
	&&\times\left[\sin\left(\frac{s\varphi}{\sqrt{3\delta(2s-1)}}\right)\right].\label{scalarpotentialsolution1}
\end{eqnarray}
Here, the dimensionless parameters $\delta$ and $s$ related to the thickness of the brane and the parameter $k$ has mass dimension one. Note that $s$ is a positive odd number. The above solution is generalizations of the thick brane discovered by Gremm et al.~\cite{Gremm:1999pj}. For $s=\delta=1$, the thick brane is returned to the solution in Ref.~\cite{Gremm:1999pj}. In addition, the solution of thick brane will change to RS-II thin brane in $\delta\rightarrow0$ limit. In this article, we focus on the case of $s>1$ and $\delta\geq1$. Plots of the warp factor~\eqref{warpfactorsolution1}, scalar field~\eqref{scalarfieldsolution1}, and scalar potential~\eqref{scalarpotentialsolution1} with different parameters are shown in Figs.~\ref{figwarpfactor},~\ref{figphi}, and~\ref{figVphi}. It can be seen that the width of the warp factor increases with the parameters $s$ and $\delta$. From Fig.~\ref{figphi}, we can see that the scalar field is a single kink for $s=1$. When $s>1$, the configuration of the scalar field becomes a double-kink, and the scalar field becomes flatter around $y=0$ as $s$ and $\delta$ increase. In addition, for the case of $s>1$, the scalar potential splits at $\phi=0$, which can be seen from Fig.~\ref{figVphi}. These phenomena indicate that the thick brane splits into two sub-branes when $s>1$, and the splitting degree deepens with the increase of $s$ and $\delta$. This can be clearly seen from the energy density distribution of the thick brane. The energy density of the above thick brane is
\begin{eqnarray}
\rho(y)&=&-3 e^{-2 H(y)} \left[A''(y)-A'(y) H'(y)+2 A'(y)^2\right]\nonumber\\
&=&\frac{-3 \delta }{y^2} \left[\left(\frac{ky}{\delta }\right)^{2 s}+1\right]^{\frac{1}{s}-2} \left(\frac{ky}{\delta }\right)^{2 s} \nonumber\\
&&\times \left[2 \delta  \left(\frac{ky}{\delta }\right)^{2 s}-2 s+1\right]. \label{energydenisty}
\end{eqnarray}
We plot the energy density in Fig.~\ref{figrho1}. It can be seen that,  the thick brane splits into two sub-branes for $s>1$, and the distance between the sub-branes increases with the $s$ and $\delta$.

In order to study the characteristic modes of the thick brane, we consider the transverse-traceless tensor perturbation of the thick brane. It is convenient to consider the fluctuation of thick brane in conformal flat coordinates $z$. Introducing the following coordinate transformation $dz=e^{H-A}dy$, the metric~(\ref{metric}) can be written as
\begin{equation}
	ds^2=e^{2A(z)}(\eta_{\mu\nu}dx^\mu dx^\nu+dz^2).
	\label{conformalmetric}
\end{equation}
Notice that $dy=dz$ when $\delta=1$. Considering the tensor perturbation of the metric of the thick brane, the perturbed metric can be written as 
\begin{eqnarray}
	g_{MN}=\left(
	\begin{array}{cc}
		e^{2A(z)}(\eta_{\mu\nu}+h_{\mu\nu}) & 0\\
		0 & e^{2A(z)}\\
	\end{array}
	\right)\label{perturbed metric},
\end{eqnarray}
where $h_{\mu\nu}$ is the transverse-traceless perturbation, which satisfies $\partial_{\mu}h^{\mu\nu}=0=\eta^{\mu\nu}h_{\mu\nu}$. Substituting Eq.~\eqref{perturbed metric} into the field equation~\eqref{field equation}, the linear equation for the tensor perturbation is
\begin{eqnarray}
	-\frac{1}{2}\Box^{(4)} h_{\mu\nu}-\frac{1}{2}h_{\mu\nu}''-\frac{3}{2}A'h_{\mu\nu}'=0, \label{mainequation}
\end{eqnarray}
where $\Box^{(4)}=\eta^{\alpha\beta}\partial_{\alpha}\partial_{\beta}$.
Introducing the following decomposition~\cite{Seahra:2005iq}
\begin{equation}
	h_{\mu\nu}=e^{-\frac{3}{2}A(z)}\Phi(t,z)e^{-i a_{j}x^{j}}\epsilon_{\mu\nu}, ~~~~\epsilon_{\mu\nu}=\text{constant},\label{decomposition1}
\end{equation}
we can obtain a one-dimensional wave equation of $\Phi(t,z)$ as
\begin{equation}
	-\partial_{t}^{2}\Phi+\partial_{z}^{2}\Phi-U(z)\Phi-a^{2}\Phi=0, \label{evolutionequation}
\end{equation}
where
\begin{eqnarray}
	U(z)=\frac{3}{2}\partial_{z}^{2}A+\frac{9}{4}(\partial_{z}A)^{2} \label{effectivepotential}
\end{eqnarray}
is the effective potential and $a$ is a constant obtained by the separation of variables. Decompose the function $\Phi(t,z)$ further into 
\begin{eqnarray}
	\Phi(t,z)=e^{-i\omega t}\phi(z).\label{decomposition2}
\end{eqnarray}
This yields a Schr\"odinger-like equation
\begin{equation}
	-\partial_{z}^{2}\phi(z)+U(z)\phi(z)=m^{2}\phi(z),\label{Schrodingerlikeequation}
\end{equation}
where $m^2=\omega^2-a^2$ denotes the mass of the KK modes. The zero mode of the above equation is
\begin{equation}
	\phi_{0}=e^{\frac{3}{2}A(z)}.\label{zeromodesolution}
\end{equation}
In addition, the Schr\"odinger-like equation~\eqref{Schrodingerlikeequation} can be rewritten as a super-symmetric form
\begin{equation}
	QQ^{\dagger}\phi(z)=m^{2}\phi(z)\label{supersymmetricform},
\end{equation}
where $Q$ and $Q^{\dagger}$ are
\begin{equation}
	Q=\partial_{z}+\frac{3}{2}\partial_{z}A	,~~~~~~~~Q^{\dagger}=-\partial_{z}+\frac{3}{2}\partial_{z}A.
\end{equation}
We can obtain another Schr\"odinger-like equation with the dual potential as
\begin{eqnarray}
	Q^{\dagger}Q\tilde{\phi}(z)=\left[-\partial_{z}^{2}+U^{\text{dual}}(z)\right]\tilde{\phi}(z)=m^{2}\tilde{\phi}(z),\label{dualSchrodingerlikeequation11}
\end{eqnarray}
where
\begin{eqnarray}
	U^{\text{dual}}(z)=-\frac{3}{2}\partial_{z}^{2}A+\frac{9}{4}(\partial_{z}A)^{2}.\label{dualeffpotentialform}
\end{eqnarray}
Interestingly, the super-symmetric quantum mechanics guarantees that the Schr\"odinger-like equations~(\ref{Schrodingerlikeequation}) and \eqref{dualSchrodingerlikeequation11} will share the same massive KK spectrum and QNMs spectrum~\cite{Cooper:1994eh,Ge:2018vjq}, and that there are no unstable tachyon mode.

\section{quasinormal modes and echo of the branes}
\label{quasinormal modes and echo of the branes}

In this section, we investigate the QNMs and the echo of the thick brane. We solve the QNMs of thick brane by the Bernstein spectral method~\cite{Fortuna:2020obg}, the direct integration method~\cite{Pani:2013pma}, and the asymptotic iteration method~\cite{Cho:2011sf}, and study the gravitational echo of thick brane by numerical evolution.

 \subsection{frequcncy domain: QNFs of thick brane}
\begin{figure}
	\subfigure[~The effective potential~\eqref{effpotentialform}]{\label{figeffectiveP}
		\includegraphics[width=0.22\textwidth]{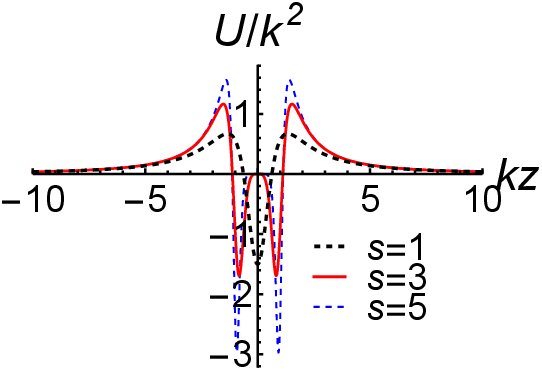}}
	\subfigure[~The dual potential~\eqref{effdpotentialform}]{\label{figdeffectiveP}
		\includegraphics[width=0.22\textwidth]{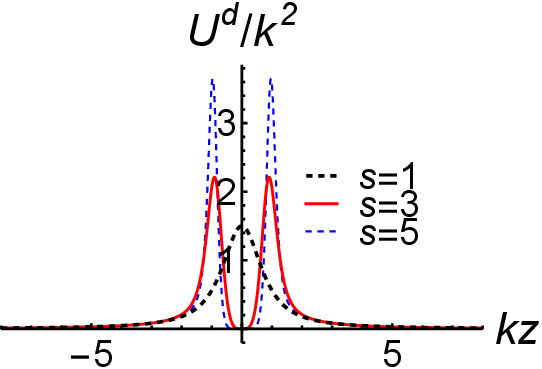}}
	\caption{The effective potential~\eqref{effpotentialform} and the dual potential~\eqref{effdpotentialform}.}\label{dandeffectiveP1}
\end{figure}

First, we use the Bernstein spectral method~\cite{Fortuna:2020obg} to solve the QNMs of the thick brane and compare the results with those obtained by the direct integration method and the asymptotic iteration method. Note that we only consider $\delta=1$, because the coordinate transformation relation between $y$ and $z$ cannot be given analytically when $\delta\neq1$. Thus, the warp factor $A(z)$, the effective potential $U(z)$, and the dual potential are given by
\begin{eqnarray}
	A(z)&=&\frac{-1}{2s}\ln\left[1+\left(kz\right)^{2s}\right],\label{zcoorwarpfactor}\\
	U(z)&=&\frac{3 k^{2s}z^{2s-2} \left(2-4s+5 k^{2s} z^{2s}\right)}{4 \left(k^{2s} z^{2s}+1\right)^2},\label{effpotentialform}\\
	U^{\text{dual}}(z)&=&\frac{3k^{2s} z^{2 s-2} \left(k^{2s}z^{2 s}+4 s-2\right)}{4 \left(k^{2s}z^{2 s}+1\right)^2}.\label{effdpotentialform}
\end{eqnarray}
We plot the above effective potential and the dual ones in Fig.~\ref{dandeffectiveP1}. We can see that, the two potentials are volcano-like and the barrier height of both potentials increases with the parameter $s$. When $s=1$, the dual potential is a pure barrier, but when $s>1$, the dual potential splits and a quasi-well appears, which can be seen from Fig.~\ref{figdeffectiveP}. The appearance of quasi-well will affect the spectrum of QNMs, and there may be a long-lived resonance mode.

We now use the Bernstein spectral method to solve the QNMs of thick brane. First, we briefly review the basic ideas of the Bernstein spectral methods. The Bernstein spectral method is a completely numerical and efficient method for solving the eigenvalues of ordinary differential equations. It has been widely applied to solve the QNMs of black holes. Consider a linear differential equation with a linear differential operator $\hat{L}(u,\omega)$
\begin{eqnarray}
\hat{L}(u,\omega)\Phi(u)=0,\label{lineareq}
\end{eqnarray}
where $\omega$ is the eigenvalue and $u\in [a,b]$ is a compact coordinate. Using the Bernstein basis, the function $\Phi(u)$ is expanded to
\begin{eqnarray}
	\Phi(u)=\sum_{k=0}^{N}C_{k}B_{k}^{N}(u),\label{basisexpand}
\end{eqnarray}
where $C_{k}$ is expansion coefficient and 
\begin{eqnarray}
B_{k}^{N}=\frac{N!}{k!(N-k)!}\frac{(u-a)^{k}(b-u)^{N-k}}{(b-a)^{N}}\label{Bernsteinbasis}
\end{eqnarray}
is the Bernstein polynomial basis. Substituting Eq.~\eqref{basisexpand} into Eq.~\eqref{lineareq} with appropriate boundary conditions, the following generalized eigenvalue equation for the set of coefficients $\textbf{C}$
 can be obtained
\begin{eqnarray}
\textbf{M}(\omega)\textbf{C}=0.\label{generalizedeq}
\end{eqnarray}
By solving the above equation~\eqref{generalizedeq}, we can get the eigenvalue of the differential equation~\eqref{basisexpand}. More details on the Bernstein spectral methods can be found in Ref.~\cite{Fortuna:2020obg}. Next, we use the Bernstein spectra method to solve the QNMs of the thick brane.

The specify form of the Schr\"odinger-like equation~\eqref{Schrodingerlikeequation} for the effective potential~\eqref{effpotentialform} is
\begin{eqnarray}
\left[	-\partial_{z}^{2}+\frac{3 k^{2s}z^{2s-2} \left(2-4s+5 k^{2s} z^{2s}\right)}{4 \left(k^{2s} z^{2s}+1\right)^2}-m^{2}\right]\phi(z)=0.\label{dualSchrodingerlikeequation} \nonumber\\
\end{eqnarray}
Since the massive KK modes escapes to infinity of extra dimension, the boundary conditions should be
\begin{equation}
	\label{boundaryconditions}
	\tilde{\phi}(z) \propto \left\{
	\begin{aligned}
		e^{im z}, &~~~~~z\to\infty.& \\
		e^{-im z},  &~~~~~z\to-\infty.&
	\end{aligned}
	\right.
\end{equation}
Because the Bernstein spectral method is effective in the compact coordinate domain, but here the extra dimensions are infinite. The Bernstein spectrum method cannot be used directly. We need to transform our coordinates to be finite. Compacting an infinite domain to a finite one by coordinate transformation is a common method in numerical relativity and in solving QNMs of black holes~\cite{Grandclement:2007sb,Cho:2011sf,Jansen:2017oag,Jaramillo:2020tuu,Chung:2023wkd}. If the form of the compact coordinates is not chosen properly, many problems will arise. Therefore, taking into account the requirements of spectral methods on the form of the equation, we introduce the transformation $u=\frac{\sqrt{4k^2 z^2+1}-1}{2k z}$. This transformation verified to be appropriate in our previous works~\cite{Tan:2022vfe,Tan:2023cra,Tan:2024url}. Now, Eq.~\eqref{dualSchrodingerlikeequation} can be rewritten as
\begin{eqnarray}
\left\{\frac{m^2}{k^2}-\frac{3 \left(\frac{u}{1-u^2}\right)^{2 (s-1)} \left[\left(\frac{u}{1-u^2}\right)^{2 s}+4 s-2\right]}{4 \left[\left(\frac{u}{1-u^2}\right)^{2 s}+1\right]^2}\right\} \phi (u)\nonumber\\
	+	\frac{\left(u^2-1\right)^3 \left(\left(u^4-1\right) \phi ''(u)+2 u \left(u^2+3\right)\phi
		'(u)\right)}{\left(u^2+1\right)^3}=0,\label{Schrodingerlikeequation1}\nonumber\\
\end{eqnarray}
where $u\in[-1,1]$. The boundary conditions~(\ref{boundaryconditions}) become
\begin{equation}
	\label{transformboundaryconditions}
	\phi(u) \propto \left\{
	\begin{aligned}
		e^{-\frac{i m/k }{2 u-2}}, &~~~ u\to 1.& \\
		e^{\frac{i m/k }{2 u+2}}, &~~~ u\to -1.&
	\end{aligned}
	\right.
\end{equation}
Further rewrite the function $\phi(u)$ as
\begin{eqnarray}
\phi(u)=\psi (u) e^{-\frac{i m/k }{2 u-2}} e^{\frac{i m/k }{2 u+2}}.\label{boundarysolutions}
\end{eqnarray}
Substituting the above expression~(\ref{boundarysolutions}) into Eq.~(\ref{Schrodingerlikeequation1}), we have a final equation that is suitable for the Bernstein spectral method
\begin{equation*}
s_{1}(u)\psi(u)+s_{2}(u)\psi'(u)+s_{3}(u)\psi''(u)=0,\label{Schrodingerlikeequation2}
\end{equation*}
where
\begin{eqnarray}
s_{1}&=&4 \left(u^2-1\right)^2 \left(u^2+1\right)\label{s1},\\
s_{2}&=&\frac{8 u \left(k \left(u^4+2 u^2-3\right)+2 i m \left(u^2+1\right)\right)}{k},\label{s2}\\
s_{3}&=&\frac{4 m^2}{k^2} \left(u^2+1\right)-\frac{8 i m}{k} \left(u^2-1\right)\nonumber\\
&&-\frac{3 \left(u^2+1\right)^3 \left[\left(\frac{u}{1-u^2}\right)^{2 s}+4 s-2\right] \left(\frac{u}{1-u^2}\right)^{2 s}}{u^2 \left[\left(\frac{u}{1-u^2}\right)^{2 s}+1\right]^2}.\label{s3}\nonumber\\
\end{eqnarray}

\begin{figure}
	\subfigure[~The first QNF]{\label{figrem1s}
		\includegraphics[width=0.22\textwidth]{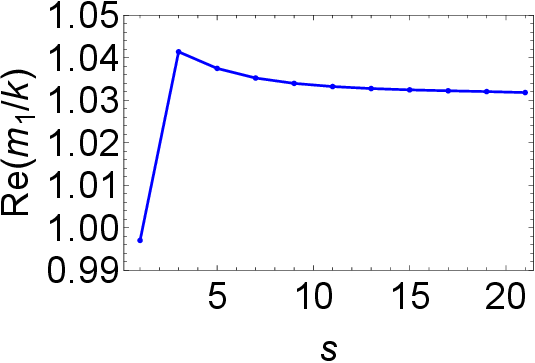}}
	\subfigure[~The first QNF]{\label{figimm1s}
		\includegraphics[width=0.22\textwidth]{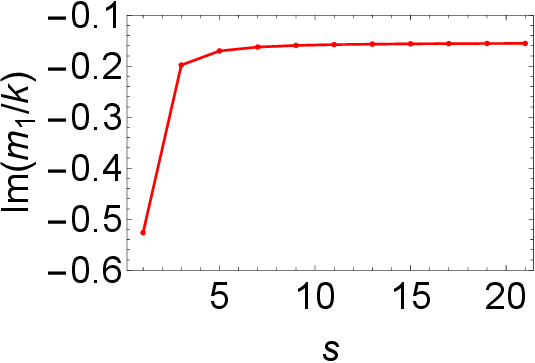}}
	\subfigure[~The second QNF]{\label{figrem2s}
		\includegraphics[width=0.22\textwidth]{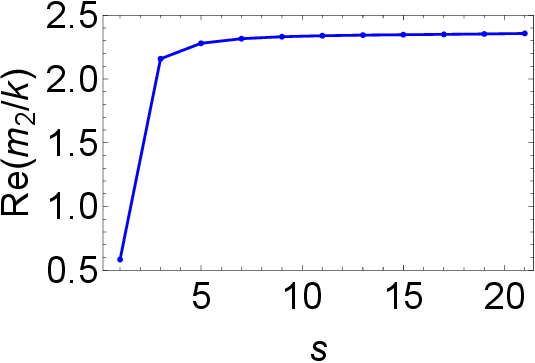}}
	\subfigure[~The second QNF]{\label{figimm2s}
		\includegraphics[width=0.22\textwidth]{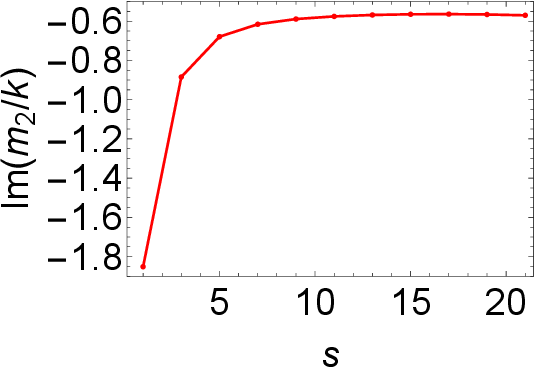}}
	\subfigure[~The third QNF]{\label{figrem3s}
		\includegraphics[width=0.22\textwidth]{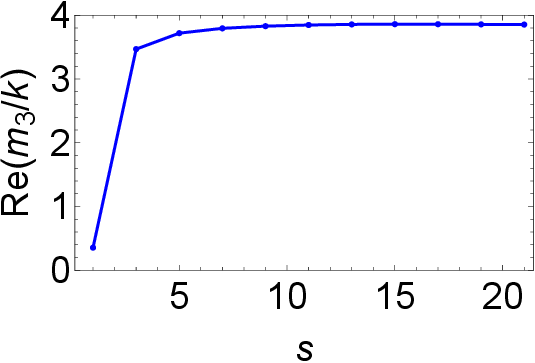}}
	\subfigure[~The third QNF]{\label{figimm3s}
		\includegraphics[width=0.22\textwidth]{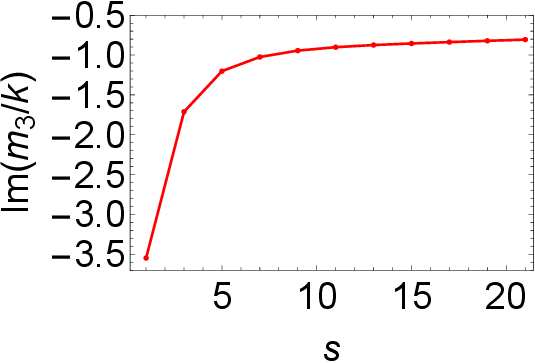}}
	\caption{The relation between the real (left panel) and the imaginary (right panel) parts of the first three QNFs and the parameter $s$.}\label{figms}
\end{figure}

\begin{figure}
	\subfigure[~$s=1$]{\label{s1dm1plot}
		\includegraphics[width=0.22\textwidth]{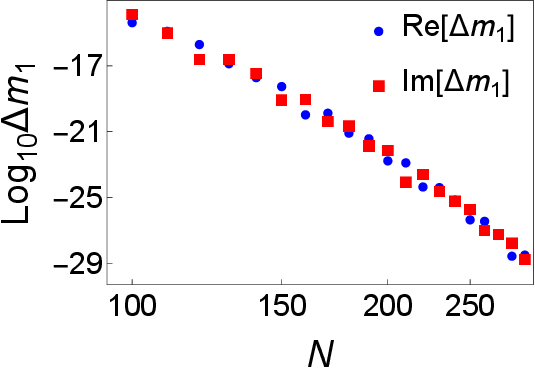}}
	\subfigure[~$s=3$]{\label{s3dm1plot}
		\includegraphics[width=0.22\textwidth]{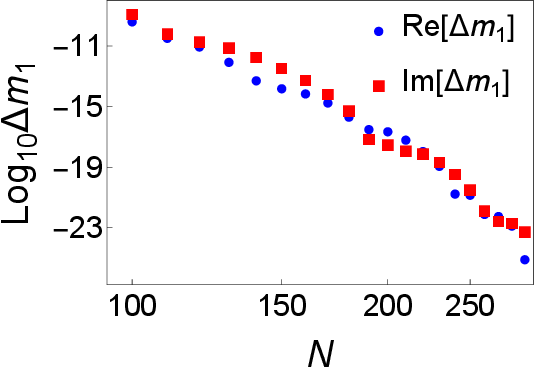}}
	\caption{The backward error of the first QNF of the thick brane depends on the spectral order $N$.}\label{figbackwarderror}
\end{figure}

We use the Mathematica package~\cite{Fortuna:2020obg} which is called SpectralB to obtain the QNMs of the thick brane. The results are compared with those obtained by the direct integration method, which are shown in Tab.~\ref{tab1}. The effect of the parameter $s$ on the first three quasinormal frequencies (QNFs) is shown in Fig.~\ref{figms}. From Tab.~\ref{tab1} and Fig.~\ref{figms}, we can see that the real parts of the second and third QNFs and the imaginary parts of the first three QNFs increase with the parameter $s$. The real part of the first QNF first increases with $s$ and then decreases. This shows that the lifetime of the first three QNMs on the brane increases with $s$. It can also be seen that when $s$ changes from 1 to 3, the first three QNFs change more dramatically. Then, with the increase of $s$, the first three QNFs change gently. This is because when $s$ from 1 to 3 the thick brane splits into two sub-brane, and the structure of the thick brane completely changes. Then $s$ increase only gradually changes the thickness of each sub-brane and the distance between sub-branes, so the quasinormal spectrum changes more gently. Moreover, to validate our numerical results by the Bernstein spectral method, we performed a backward error analysis on the results. The backward error is defined as~\cite{Chung:2023wkd}
	\begin{equation}
\Delta m_{n}=|m_{n}(N+i)-m_{n}(N)|.\label{backward error}
	\end{equation}
We analyze the backward error of the first QNF for $s=1$ and $s=3$, and the results are shown in Fig.~\ref{figbackwarderror}. The results show that the truncation error of the first QNF decreases exponentially with the spectral order $N$, consistent with the exponential convergence predicted by the spectral method.

\begin{table*}[htbp]
	\begin{tabular}{|c|c|c|c|c|}
		\hline
		$\;\;s\;\;$  &
		$\;\;n\;\;$  &
		$\;\;\text{Bernstein spectral method}\;\;$  &
		$\;\;\;\;\;\;\;\;\text{Direct integration method}\;\;\;\;\;\;\;$ 	&
		$\;\;\;\;\;\;\;\;\text{Time evolution}\;\;\;\;\;\;\;$\\
		\hline
		~  &~   &~~~~$\text{Re}(m/k)$  ~~  $\text{Im}(m/k)~~$  &$~~~~~~\text{Re}(m/k)$ ~~ $\text{Im}(m/k)~~$  &$~~~~~~\text{Re}(m/k)$ ~~ $\text{Im}(m/k)~~$     \\
		1    &1   &0.99702~~ -0.52636           &~~0.99702 ~~ -0.52636               &~~0.99117~~~~   -0.52755\\
		3    &1   &1.04138~~ -0.19783           &~~1.04138 ~~ -0.19783               &~~1.04012~~~~   -0.19778\\
	       	 &2   &2.15785~~ -0.88434           &~~2.15785 ~~ -0.88434               &~~ non~~~~~~~~non \\
		5    &1   &1.03749~~ -0.17009           &~~1.03749 ~~ -0.17009               &~~1.03666~~~~   -0.17000\\
	         &2   &2.27955~~ -0.67896           &~~2.27955 ~~ -0.67896               &~~  non~~~~~~~~non \\
     	7    &1   &1.03521~~ -0.16249           &~~1.03521 ~~ -0.16249               &~~1.03449~~~~   -0.16240\\
             &2   &2.31570~~ -0.61571           &~~2.31570 ~~ -0.61571               &~~  non~~~~~~~~non \\		
	    9    &1   &1.03395~~ -0.15933           &~~1.03395 ~~ -0.15933               &~~1.03327~~~~   -0.15923\\	
	         &2   &2.33110~~ -0.58903           &~~2.33110 ~~ -0.58901               &~~  non~~~~~~~~non\\
		11   &1   &1.03321~~ -0.15770           &~~1.03321 ~~ -0.15770               &~~1.03254~~~~   -0.15760\\
	         &2   &2.33898~~ -0.57558           &~~2.33896 ~~ -0.57539               &~~  non~~~~~~~~non \\
		13   &1   &1.03274~~ -0.15676            &~~1.03273 ~~ -0.15676               &~~1.03208~~~~   -0.15666\\
	         &2   &2.34363~~ -0.56831           &~~2.34347 ~~ -0.56754               &~~  non~~~~~~~~non\\
		\hline
	\end{tabular}
	\caption{The frequencies of low overtone modes with different $s$ by the Bernstein spectral method, the direct integration method, and the time evolution.\label{tab1}}
\end{table*}

\begin{figure}
	\subfigure[~$s=1$]{\label{figqnms1plotten}
		\includegraphics[width=0.22\textwidth]{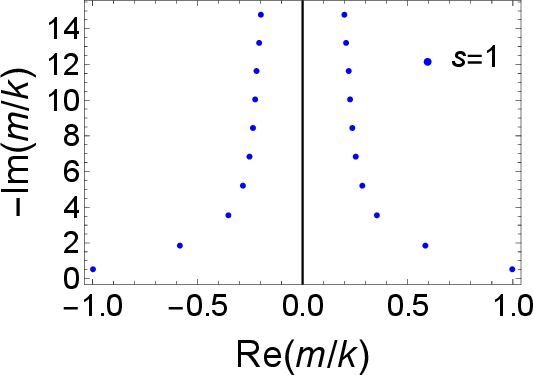}}
	\subfigure[~$s>1$]{\label{figqnms31321plotten}
		\includegraphics[width=0.22\textwidth]{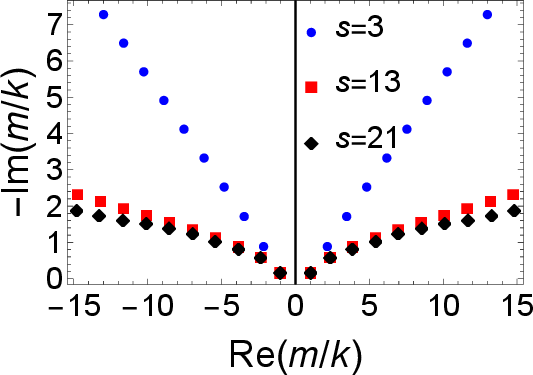}}
	\caption{The first ten quasinormal frequencies of the brane solved by the Bernstein spectral method with different $s$.}\label{qnmplotten}
\end{figure}

In addition, we also investigated the high overtone modes of the thick brane. The results are shown in Tabs.~\ref{tab2},~\ref{tab3}, and Fig.~\ref{qnmplotten}. From Fig.~\ref{figqnms1plotten} we can see that the real part of the first ten QNFs decreases with the overtone number $n$. This is very similar to the Schwarzschild black hole situation. Because the dual potential of the thick brane when $s=1$ is similar to the shape of the effective potential of the Schwarzschild black hole. But when the brane splits, the situation is very different. Table~\ref{tab3} and Fig.~\ref{figqnms31321plotten} show the first ten QNFs of thick brane with different parameters $s$. From Fig.~\ref{figqnms31321plotten} we can clearly see that, in contrast to the case of $s=1$ (the brane does not split), the real part of the first ten QNFs increases with $n$. Therefore, the splitting of thick brane has a great influence on its quasinormal spectrum. 

On the other hand, these results also demonstrate the stability of low overtone modes. As can be seen from Fig.~\ref{figqnms31321plotten}, when the parameter $s$ changes, the modes with small absolute value of imaginary part, especially the modes with the smallest imaginary part (the fundamental mode), changes little. On the contrary, higher overtone modes vary more. This is the same as the result indicated recently by the black hole pseudo-spectrum~\cite{Jaramillo:2020tuu}, that is, the high overtone modes are highly sensitive to changes in the effective potential, while the fundamental mode is stable. In general, the results in the above frequency-domain show that the quasinormal spectrum of the split thick brane is completely different from that of the unsplit thick brane. In order to study the characteristic mode of the split thick brane more completely, we will perform the time-domain analysis in the next section.

\begin{table*}[htbp]
	\begin{tabular}{|c|c|c|c|}
		\hline
		$\;\;s\;\;$  &
		$\;\;n\;\;$  &
		$\;\;\text{Bernstein spectral method}\;\;$  &
		$\;\;\;\;\;\;\;\;\text{Asymptotic iteration method}\;\;\;\;\;\;\;$ 	\\
		\hline
		~  &~   &~~~~$\text{Re}(m/k)$  ~~  $\text{Im}(m/k)~~$  &$~~~~~~\text{Re}(m/k)$ ~~ $\text{Im}(m/k)~~$   \\
		~    &1   &0.99702~~ -0.52636           &~~0.99702 ~~  -0.52636                \\
		~    &2   &0.58371~~ -1.85156           &~~0.58371 ~~ -1.85156           \\
		~  	 &3   &0.35314~~ -3.54785           &~~0.35271 ~~ -3.54771           \\
		1    &4   &0.28388~~ -5.20906           &~~0.28290 ~~ -5.20962         \\
		~  	 &5   &0.25268~~ -6.83448           &~~0.24770 ~~ -6.83235              \\
		~    &6   &0.23631~~  -8.44131          &~~0.23933 ~~ -8.43357               \\
		~	 &7   &~0.22560~~ -10.04081         &~~~0.21571~~ -10.05330              \\		
		~    &8   &~0.21927~~ -11.63293         &~~~0.21294~~ -11.64998              \\	
		\hline
	\end{tabular}
	\caption{The first eight QNFs using the Bernstein spectral method and the asymptotic iteration method.\label{tab2}}
\end{table*}

\begin{table*}[htbp]
	\begin{tabular}{|c|c|c|c|}
		\hline
		$\;\;s\;\;$  &
		$\;\;n\;\;$  &
		$\;\;\text{Bernstein spectral method}\;\;$  &
		$\;\;\;\;\;\;\;\;\text{Direct integration method}\;\;\;\;\;\;\;$ 	\\
		\hline
		~  &~   &~~~~$\text{Re}(m/k)$  ~~  $\text{Im}(m/k)~~$  &$~~~~~~\text{Re}(m/k)$ ~~ $\text{Im}(m/k)~~$   \\
		~    &1   &1.03203~~ -0.15550           &~~1.03203 ~~  -0.15547                \\
		~    &2   &2.35270~~ -0.56578           &~~2.34944 ~~ -0.55697           \\
	    ~    &3   &3.85612~~ -0.82156           &~~3.87348 ~~ -0.84437            \\
		~    &4   &5.43447~~ -1.03258           &~~5.42952 ~~ -1.05052         \\
	    19   &5   &6.96103~~ -1.23800           &~~6.99144 ~~ -1.22080               \\
		~    &6   &8.54970~~ -1.39227           &~~8.55497 ~~ -1.37331               \\
	    ~    &7   &10.09836~~ -1.53539          &~~10.11913~~ -1.51628              \\		
		~    &8   &11.67300~~ -1.64354          &~~11.68364~~ -1.65388              \\	
	    ~    &9   &13.27121~~ -1.78436          &~~13.24841~~ -1.78836            \\
		~    &10  &14.79788~~ -1.92809          &~~14.81335~~ -1.92101                     \\
							\hline
		~    &1   &1.03180~~   -0.15534        &~~1.03191~~  -0.15526               \\
		~    &2   &2.35691~~ -0.57003          &~~2.35037~~  -0.55528           \\
		~  	 &3   &3.85274~~ -0.80636          &~~3.87600~~  -0.83896            \\
		~    &4   &5.43948~~ -1.01395          &~~5.43365~~  -1.03948          \\
		21 	 &5   &6.95075~~ -1.22743          &~~6.99710~~  -1.20254                \\
		~    &6   &8.55633~~ -1.37537          &~~8.56201~~  -1.34654               \\
		~	 &7   &10.09980~~ -1.50853         &~~10.12741~~ -1.47998              \\		
		~    &8   &11.67796~~ -1.59551         &~~11.69306~~ -1.60725              \\	
		~	 &9   &13.28684~~ -1.72391         &~~13.25887~~ -1.73083            \\
		~    &10  &14.80613~~ -1.86478         &~~14.82482~~ -1.85214                     \\
		\hline
	\end{tabular}
	\caption{The first ten QNFs using the Bernstein spectral method and the direct integration method.\label{tab3}}
\end{table*}

\subsection{Time domain: quasinormal and echo waveforms}

\begin{figure}
	\subfigure[~$kz_\text{ext} = 0$]{\label{qnmsplotevenz0}
		\includegraphics[width=0.22\textwidth]{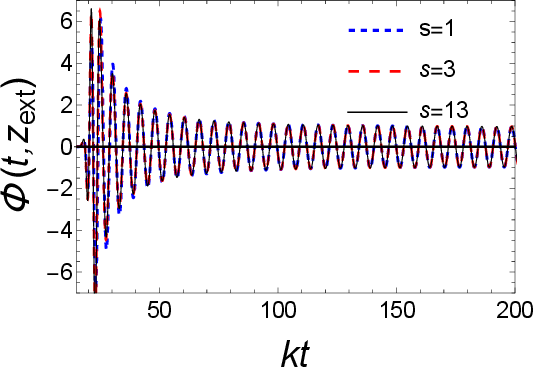}}
	\subfigure[~$kz_\text{ext} = 5$]{\label{qnmsplotevenz5}
	\includegraphics[width=0.22\textwidth]{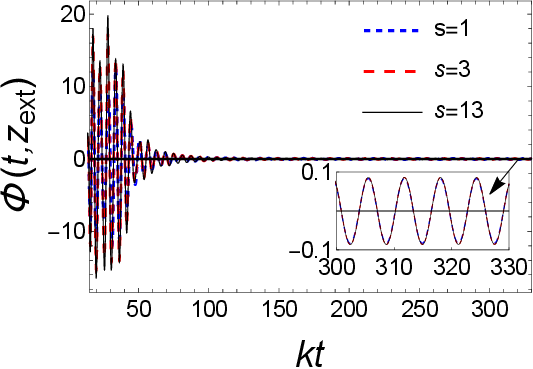}}		
		\subfigure[~$kz_\text{ext} =10$]{\label{qnmsplotevenz10}
		\includegraphics[width=0.22\textwidth]{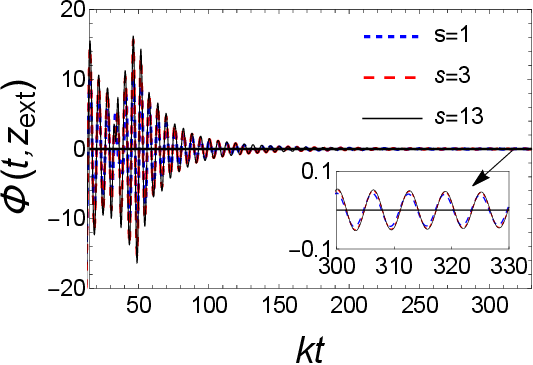}}		
			\caption{Evolution waveforms of the Gauss packets~\eqref{gausspulseinitialwavepackage} at different locations. The data are extracted at the points $kz_\text{ext} = 0, 5, 10$.}\label{qnmsploteven}
\end{figure}

\begin{figure}[htbp]
	\centering
	\includegraphics[width=0.4\textwidth]{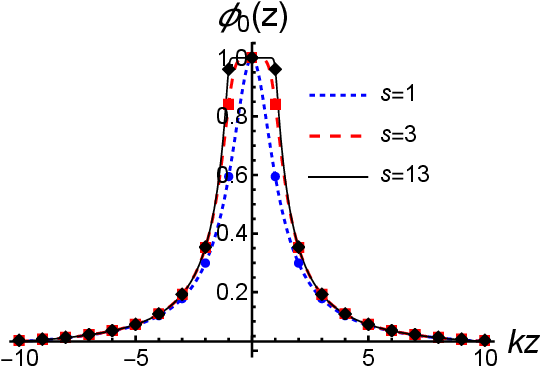}
	\caption{Compare the zero mode (dot) exited by the Gauss pulse with the analytical zero mode~\eqref{zeromodesolution} (line).}\label{zeromoderfit}
\end{figure}

We have studied the QNMs of the thick brane in the frequency-domain by various methods. These results show how these fluctuations oscillate and decay when the thick brane is perturbed. But this is not complete, only in the frequency-domain research lacks a description of the evolution of these fields waveform, the relative amplitude, etc. Therefore, it is necessary to consider the evolution of the $1+1$ wave equation~\eqref{evolutionequation}. First, we still consider the case of $\delta=1$, where the effective potential has an analytical expression in $z$ coordinate. We can easily change the coordinates to the light-cone coordinates $u=t+z$ and $v= t-z$ for evolution. Then evolution equation~\eqref{evolutionequation} can be written as
\begin{eqnarray}
	\left(4\frac{\partial^{2}}{\partial u\partial v}+U+a^{2}\right)\Phi(u,v)=0. \label{uvevolutionequation}
\end{eqnarray}
We consider firstly the initial data is a Gauss packet with width $\sigma$ and located at $v_{c}$
\begin{eqnarray}
	\Phi(0,v)=e^{\frac{-(v-v_{c})^{2}}{2\sigma^{2}}}, ~~~\Phi(u,0)=e^{\frac{-v_{c}^{2}}{2\sigma^{2}}},\label{gausspulseinitialwavepackage}
\end{eqnarray}
We chose the Gaussian wave packet with width $k\sigma=1$ and located at $kv_{c}=20$. We set the parameter $a$ to $a/k=1$. The evolution waveform of Gauss pulse is shown in Fig.~\ref{qnmsploteven}, which show how the initial wave packet evolves over time at different locations in the extra dimension. It can be seen that after the initial oscillation damping, the evolution of the wave packet tends to be a sinusoid. The frequencies of these final sinusoids does not depend on the extraction point $z_{\text{ext}}$, but the amplitude decreases with $z_{\text{ext}}$. This is due to the existence of a bound zero mode of the thick brane, which does not decay. Specifically, according to the analytic solution ~\eqref{zeromodesolution} of the zero mode and the decomposition ~\eqref{decomposition2}, the time-dependent solution of the zero mode is $\phi_{0}(t,z)=\sin(\omega t)\left[1+\left(kz\right)^{2s}\right]^{\frac{-3}{4s}}$. Thus, if we observe at the extra dimensional fixed point, the evolutionary behavior of the zero mode with time is the undecayed sinusoid. To confirm our analysis, we extract the late maximum amplitude of the evolved waveform at different points and compare it with the analytic zero mode~\eqref{zeromodesolution}, the results are shown in Fig.~\ref{zeromoderfit}. It can be seen that the profile obtained by the numerical evolution is consistent with the shape of the analytic zero mode. This confirms our previous analysis and strengthens the confidence of our numerical evolutionary results.

\begin{figure}
		\centering
		\includegraphics[width=0.4\textwidth]{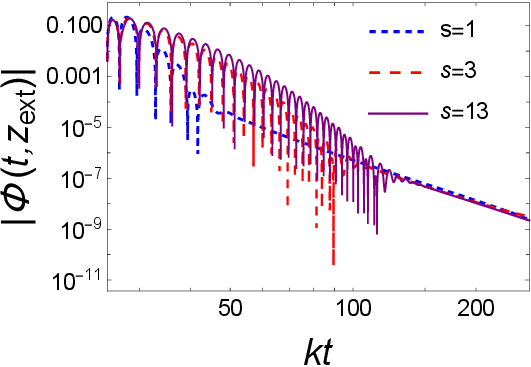}
	\caption{Evolution waveform of the static odd packet~\eqref{oddinitialwavepackage} at $kz_{\text{ext}}=30$ in logarithmic scale.}\label{qnmsplotodd1}
\end{figure}

However, because the information of the QNMs is covered by the zero mode, it is difficult to obtain the information of the QNMs of the thick brane from the evolution of the Gaussian wave packet. There are two ways to solve this problem, one is to place the extraction point far away from the zero point of the extra dimension $z$, then the effect of the zero mode becomes very weak. But that requires a lot of computing resources. Another method is given by the symmetry of the brane. Because of the $Z_2$ symmetry, the KK mode can only be odd or even. It is clear that the zero mode is an even function, and the first excited state, the first QNM, is odd-parity. Therefore, when we consider that the initial wave packet is static and odd-parity, only the odd-parity QNMs will be excited, so that we can study the evolution waveform of odd-parity QNMs. The form of a static odd wave packet is as follows
\begin{eqnarray}
	\Phi(0,v)=\sin\left(\frac{kv}{2}\right)e^{\frac{-k^2v^{2}}{4}}, \nn\\
	\Phi(u,0)=\sin\left(\frac{ku}{2}\right)e^{\frac{-k^2u^{2}}{4}}.\label{oddinitialwavepackage}
\end{eqnarray} 
We plot the waveforms for different $s$ in Fig.~\ref{qnmsplotodd1}. Note that we set $a=0$ for this case. It can be seen that, after the initial burst phase, the wave packet evolution has the characteristic of a typical QNM: the wave packet first exponentially decay, and then appears a power-law damping tail at a late stage. It can also be seen from Fig.~\ref{qnmsplotodd1} that with the increase of $s$, the damping of the initial wave packet slows down, which is consistent with the results obtained in the frequency domain. In fact, the frequency of the first QNM can be obtained by fitting the evolutionary data. Table~\ref{tab1} also lists the frequency of the first QNM of the thick brane obtained by numerical evolution for different parameters $s$, which is consistent with the results obtained in the frequency domain. These results show that, for thick brane, there are discrete characteristic modes in continuous massive KK modes. The properties of these characteristic modes depend on the configuration of the extra dimensions.

\begin{figure}
	\subfigure[~The effective potential $U(z)$]{\label{effectivePd}
		\includegraphics[width=0.22\textwidth]{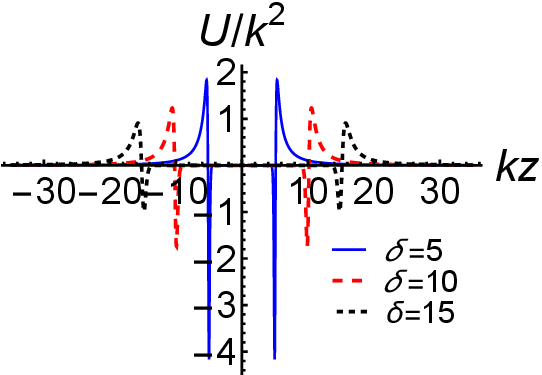}}
	\subfigure[~the energy density $\rho(z)$]{\label{rho123Pd}
		\includegraphics[width=0.22\textwidth]{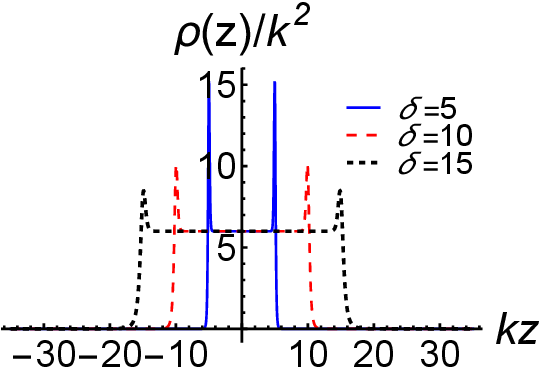}}
	\caption{The effective potential and the energy density for different $\delta$ with $s=35$.}\label{urhoecho}
\end{figure}

Now, let us turn our attention to another interesting phenomenon: the echo. When $s>1$, the thick brane splits, and the dual potential results in a quasi-well. The quasi-well is similar to a resonant cavity; thus, the waves will be quasi-bound in the cavity and slowly leak out, resulting in an echo. In the previous part, we did not observe the echo because the two sub-barriers of the effective potential, or the two sub-branes of the thick brane, are too close to support the presence of the echo signal. But when we adjust $\delta$, things change.
We plot the shapes of the effective potential and energy density corresponding to different $\delta$ with $s=35$ in Fig.~\ref{urhoecho}. It can be seen that the distance between the two sub-barriers of the effective potential and the peaks of the energy density increases with $\delta$, indicating that the degree of brane splitting intensifies. 

\begin{figure*}
	\subfigure[~$s=35$, $\delta=5,kv_{c}=5$]{\label{figd5s5z07psitz0}
		\includegraphics[width=0.45\textwidth]{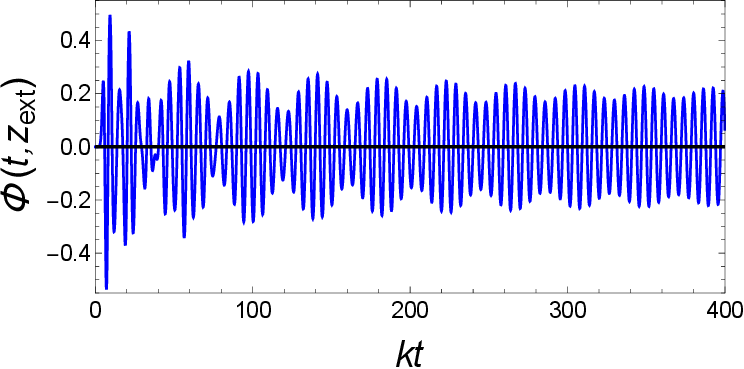}}
	\subfigure[~$s=35$, $\delta=10,kv_{c}=11$]{\label{figd10s35z011psitz0}
		\includegraphics[width=0.45\textwidth]{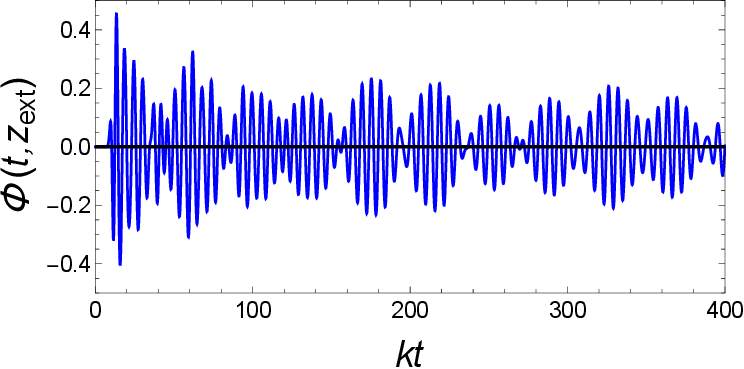}}	
	\subfigure[~$s=35$, $\delta=15,kv_{c}=16$]{\label{figd15s35z016psitz0}
		\includegraphics[width=0.45\textwidth]{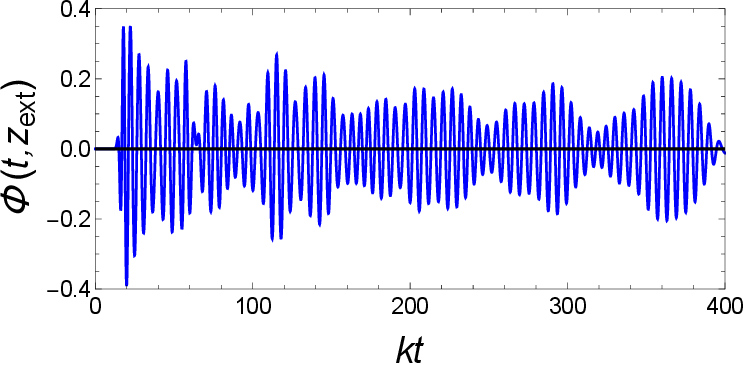}}
	\caption{Evolution waveform of the Gauss packet~\eqref{gausspulseinitialwavepackage} at $kz_{\text{ext}}=0$ for different $\delta$. The parameter of the Gaussian wave packet is $k\sigma=1$.}\label{waveformeven1}
\end{figure*}

We now consider the evolution of Gaussian wave packets on these thick branes. We focus on the Gaussian wave packet incident to the left from the right peak of the energy density to simulate the emission of gravitational waves from one sub-brane to another. Figure~\ref{waveformeven1} shows the evolution waveform of the incident Gaussian wave packet on the brane over time at different $\delta$ observed at the zero point of the extra dimension. We can see that, for the parameter we selected $\delta$, the echo signal appears in the time-domain evolution waveform. When $\delta$ is small, the echo signal is masked by the zero mode after some time, as shown in Fig.~\ref{figd5s5z07psitz0}. As $\delta$ increases further, the echo signal becomes apparent, but the overall evolutionary waveform seems to become less regular. This happens in the evolution of a massive field in a black hole background and is known as the beating effect~\cite{Witek:2012tr}. This phenomenon can be explained by the beat modulation between two or more long-lived modes with similar frequencies. This beat modulation depends on the relative intensity of the different overtones, which in turn depends on the extraction point, as in the case of guitar strings, a specific mode cannot be excited at its nodes. These characteristics should be universal as long as there are at least two long-lived modes present.

\begin{figure*}		
	\subfigure[~$s=35$, $\delta=10$, $kz_\text{ext}=10$]{\label{figd10s35z011psitz10}
		\includegraphics[width=0.45\textwidth]{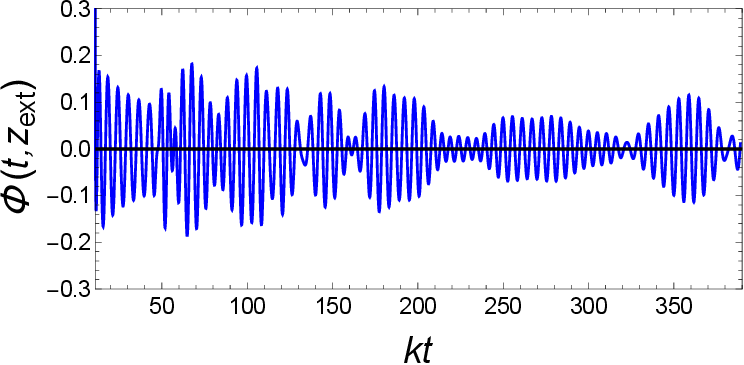}}		
	\subfigure[~$s=35$, $\delta=10$,$kz_\text{ext}=10$]{\label{figd10s35z011psitz10fplot}
		\includegraphics[width=0.45\textwidth]{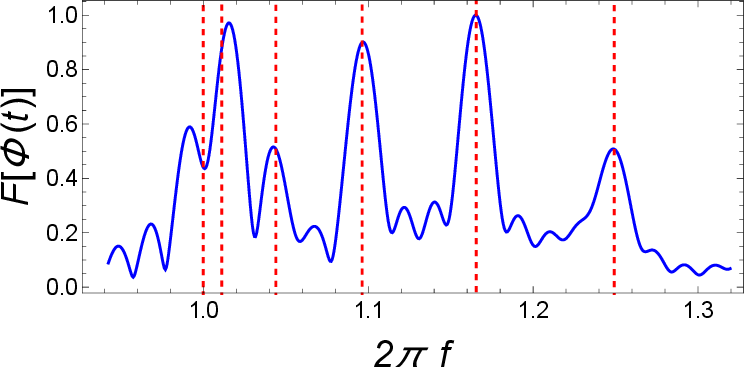}}
	\subfigure[~$s=35$, $\delta=10$, $kz_\text{ext}=-10$]{\label{figd10s35z011psitzn10}
		\includegraphics[width=0.45\textwidth]{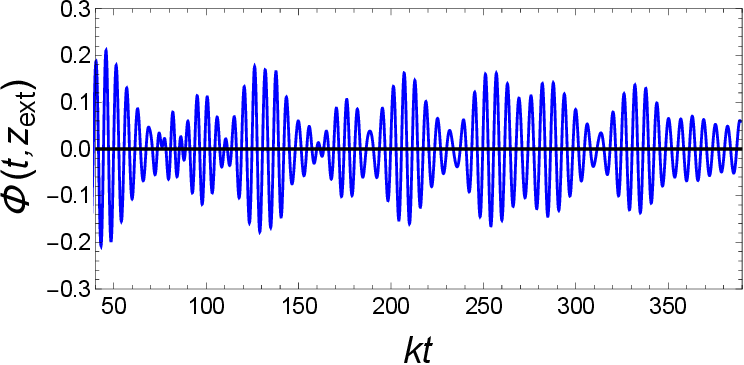}}
	\subfigure[~$s=35$, $\delta=10$, $kz_\text{ext}=-10$]{\label{figd10s35z011psitzn10fplot}
		\includegraphics[width=0.45\textwidth]{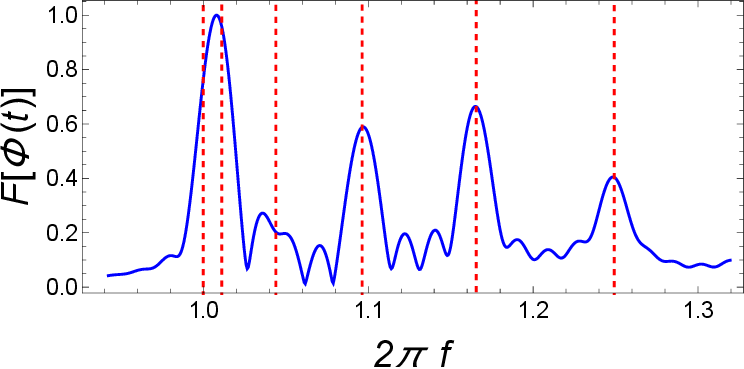}}				
	\subfigure[~$s=35$, $\delta=10$, $kz_\text{ext}=50$]{\label{figd10s35z011psitz50}
		\includegraphics[width=0.45\textwidth]{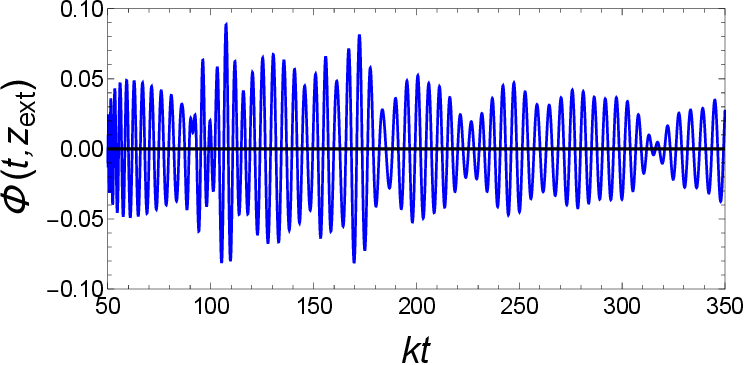}}
	\subfigure[~$s=35$, $\delta=10$, $kz_\text{ext}=50$]{\label{figd10s35z011psitz50fplot}
		\includegraphics[width=0.45\textwidth]{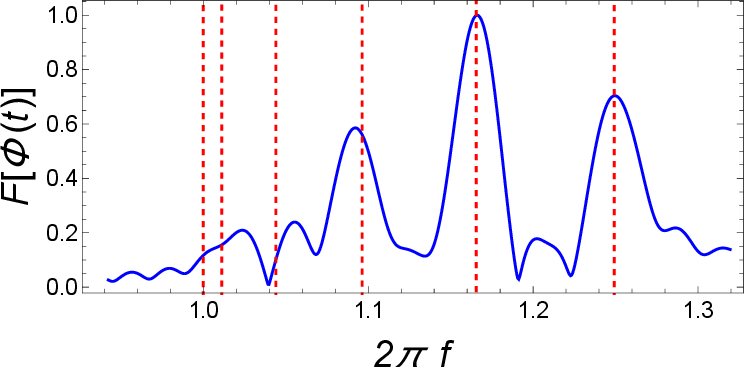}}						
	\caption{Left panel: Evolution waveform of the Gauss packet~\eqref{gausspulseinitialwavepackage} at different locations $kz_{\text{ext}}$. Right panel: Corresponding spectrum obtained by the Fourier transformation. The red dashed vertical lines correspond from left to right to the oscillation frequency of the zero mode, first QNM, second QNM, and so on. The parameters of Gauss packet~\eqref{gausspulseinitialwavepackage} are $kv_c=11$ and $k\sigma=1$.}\label{waveformeven2}
\end{figure*}

To understand the beating effect more directly, it is useful to observe the waveform evolution of different extraction points and the corresponding Fourier spectrum. In Fig.~\ref{waveformeven2}, we show the waveforms and corresponding Fourier spectrum of Gaussian wave packets evolving over time at different extraction points when $\delta=10$ and $s=35$. The extraction points are selected as $kz_{\text{ext}}=10$ (the location of the sub-brane where the gravitational wave occurs), $kz_{\text{ext}}=-10$ (the location of another sub-brane) and $kz_{\text{ext}}=50$ (the simulated infinity of extra dimensions). These results clearly show that several peaks correspond to different mode contributions of KK gravitons (zero mode or long-lived quasi-normal modes), and the relative amplitudes of these peaks vary significantly as the extraction point changes. This confirms our previous analysis of the beating effect.

\begin{figure*}
	\subfigure[~$s=35$, $\delta=10$, $a/k=1/2$]{\label{figd10s35z011psitz0a05}
		\includegraphics[width=0.45\textwidth]{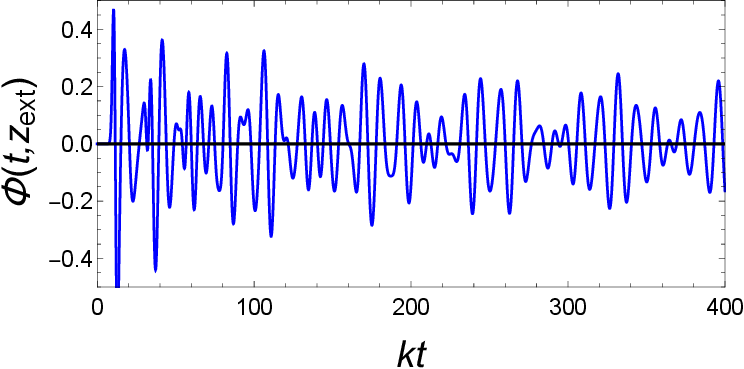}}
	\subfigure[~$s=35$, $\delta=10$, $a/k=3/2$]{\label{figd10s35z011psitz0a105}
		\includegraphics[width=0.45\textwidth]{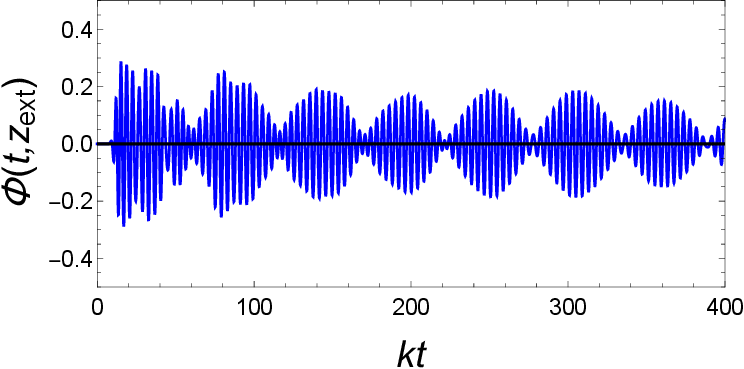}}
	\caption{Evolution waveforms of the Gauss packet~\eqref{gausspulseinitialwavepackage} at $kz_{\text{ext}}=0$ for different $a$. The parameters of the Gauss packet~\eqref{gausspulseinitialwavepackage} are set to $kv_c=11$ and $k\sigma=1$.}\label{waveformeven3}
\end{figure*}

In addition, we also briefly consider the effect of the parameter $a$, the results are shown in Fig.~\ref{waveformeven3}. It can be seen that the oscillation frequency of the evolving waveform increases with $a$. This is due to the oscillation frequency $\omega^{2}=a^{2}+m^{2}$. The above numerical evolution results show that when the degree of brane splitting is high, the oscillation of the massive KK mode on the brane will indeed echo. However, the existence of zero mode and beating effect masks a lot of information, and it is difficult to obtain information such as the period of the echo.

\begin{figure*}
	\subfigure[~$s=35,\delta=5$]{\label{oddfigd5s35psitz5}
		\includegraphics[width=0.45\textwidth]{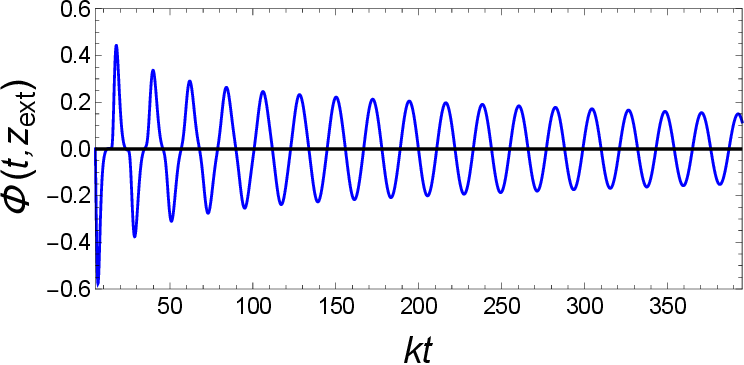}}
	\subfigure[~$s=35,\delta=5$]{\label{logoddfigd5s35psitz5}
		\includegraphics[width=0.45\textwidth]{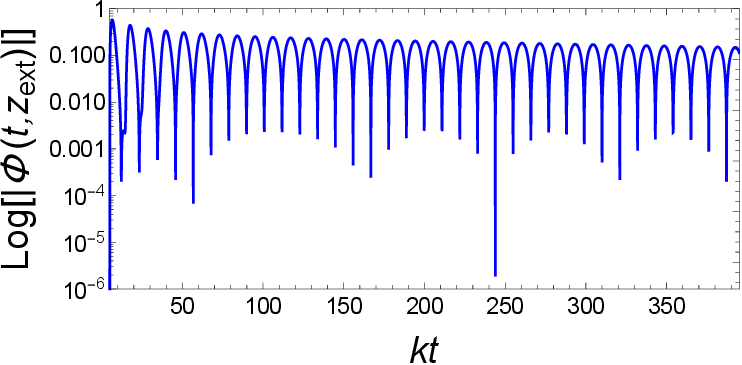}}		
	\subfigure[~$s=35, \delta=15$]{\label{oddfigd15s35psitz5}
		\includegraphics[width=0.45\textwidth]{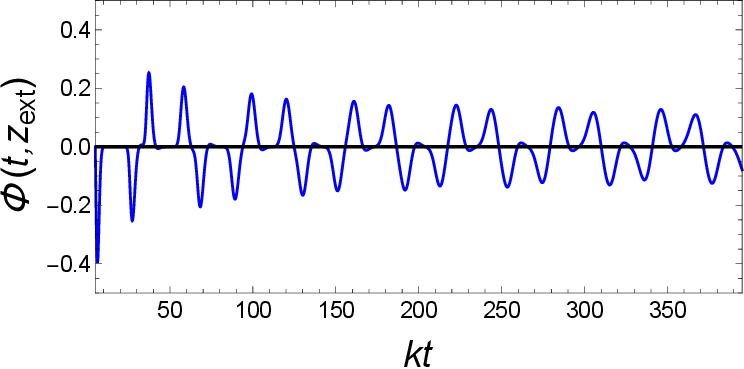}}
	\subfigure[~$s=35, \delta=15$]{\label{logoddfigd15s35psitz5}
		\includegraphics[width=0.45\textwidth]{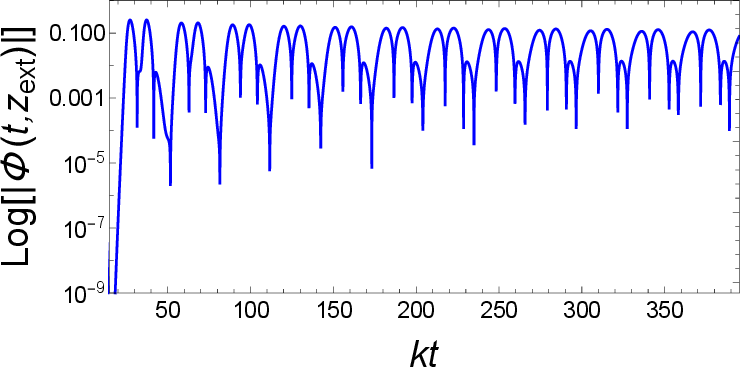}}
	\subfigure[~$s=35, \delta=25$]{\label{oddfigd25s35psitz0}
		\includegraphics[width=0.45\textwidth]{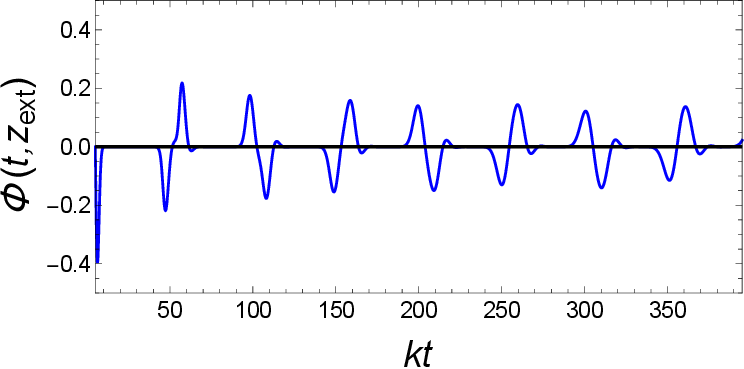}}
	\subfigure[~$s=35, \delta=25$]{\label{logoddfigd25s35psitz0}
		\includegraphics[width=0.45\textwidth]{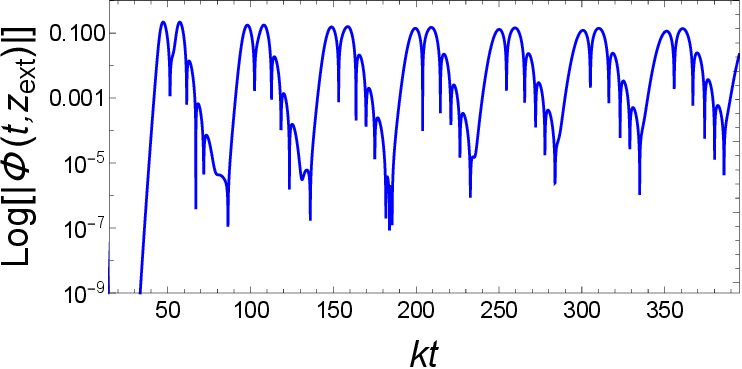}}
	\caption{Left panel: Evolution waveforms of the static odd-parity wave packet~\eqref{oddinitialwavepackage} at $kz_{\text{ext}}=5$ for different values of $\delta$. Right panel: Same as the left panel but plotted on a semi-log scale.}\label{waveformodd1}
\end{figure*}

 Thus, we consider the evolution of the static odd-parity wave packet that excludes the effects of zero mode. The wave packet is also chosen in the form of Eq.~\eqref{oddinitialwavepackage}. Figure~\ref{waveformodd1} shows the evolution of odd-parity wave packets with different $\delta$. We can see that when $\delta$ is large, the wave packet oscillates in the quasi-well to produce an echo, and the amplitude of the echo decreases with time. The period of the echo can be clearly observed in the corresponding semi-log coordinate. It can be seen that when $\delta=15$, the echo period is about 30, and when $\delta=25$, the echo period is about 50, which is exactly the same as the width of the corresponding potential well.

\begin{figure*}
	\subfigure[~$s=35$, $\delta=15$, $kz_{\text{ext}}=15$]{\label{oddfigd15s35psitz15}
		\includegraphics[width=0.45\textwidth]{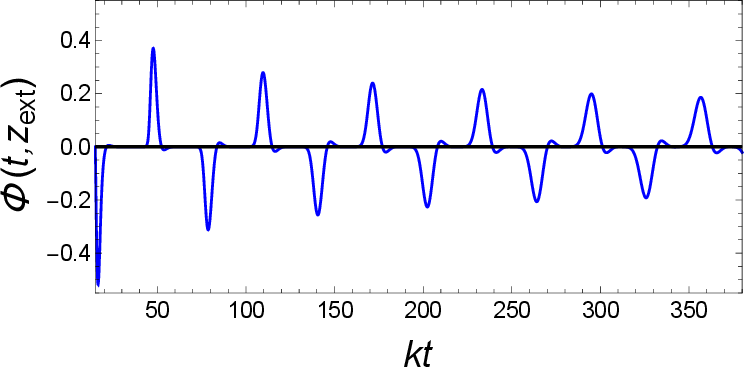}}	
	\subfigure[~$s=35$, $\delta=15$, $kz_{\text{ext}}=50$]{\label{oddfigd15s35psitz50}
		\includegraphics[width=0.45\textwidth]{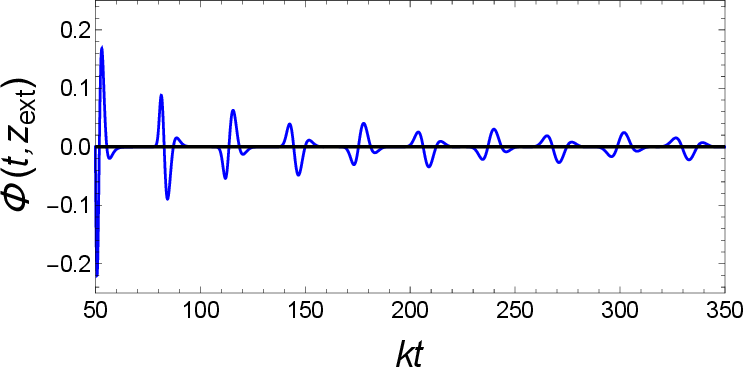}}
	\caption{Evolution waveforms of the static odd packet~\eqref{oddinitialwavepackage} at different points $kz_{\text{ext}}$.}\label{waveformodd2}
\end{figure*}

On the other hand, in the case of odd-parity initial data, we do not seem to observe a beating phenomenon. To verify this, we select different extraction points to observe the evolution waveform, and the results are shown in Fig.~\ref{waveformodd2}. It can be seen that the shape of waveforms at different extraction points is basically the same, which indicates that there is indeed no beating effect under the odd-parity initial data when $a=0$. This may be because the odd-parity wave packet only excites the odd QNMs, so it does not interfere with the zero mode. Additionally, the frequency gap between the different odd QNMs is large enough that they do not affect each other.

According to the above results, the split thick brane model exhibits a rich phenomenology, including changes in the spectral structure of QNMs, the emergence of long-lived QNMs, echoes, and the beating effect. We also consider how the thick brane affects the scales involved in gravitational wave detection. According to Eq.~\eqref{mainequation}, the equation of motion for $h_{\mu\nu}$ on the brane is given by:
\begin{equation}
\square^{(4)} h^{n}_{\mu\nu}-m_{n}^{2}h^{n}_{\mu\nu}=0,\label{fourdimensionaleq}
	\end{equation}
where $m_{n}^{2}=\omega^{2}-a^{2}$ is the mass of each KK mode. The zero mode ($n=0$) corresponds to massless gravitons, while higher modes ($n>0$) represent massive spin-2 mode towers.

When an event (such as a black hole merger) produces gravitational waves, it excites $h_{\mu\nu}(x^{\mu};z)$, which decomposes into infinite mode towers on the brane. The zero mode, with $m_{0}=0$, corresponds to gravitational waves with frequencies associated with the source event. The higher modes, with $m_{n}>0$, have frequencies given by $\omega_{n}=m_{n}$, neglecting the effect of $a$. If the KK mode masses are high enough, their frequencies will lie outside the range covered by current gravitational wave detectors, making only the zero mode accessible. However, if the KK mode masses are sufficiently low, their frequencies may fall within the range of current or future experiments. In this case, one might hope to detect a characteristic signature of extra dimensions: a tower of modes with a constant frequency gap $\Delta \omega=m_{1}$. In Tab.~\ref{tab4}, we list the frequency ranges and corresponding $m_1$ of current or under construction gravitational wave detectors such as Lisa~\cite{Lisa:2017}, LIGO-Virgo~\cite{LIGOScientific:2016aoc}, and The CERN Axion Solar Telescope (CAST)~\cite{Ejlli:2019bqj,Aggarwal:2020olq}.
\begin{table}[htbp]
	\begin{tabular}{|c|c|c|}
		\hline
		Detector  &
		$f_{\text{GW}}$(Hz) &
		$m_{1}$(eV) 	\\
		\hline
		LISA  &~$10^{-4}-10^{0}$   &  $10^{-31}-10^{-27}$   \\
		\hline
		LIGO-Virgo  &~$10^{1}-10^{4}$   & $10^{-26}-10^{-23}$    \\
		\hline
		CAST &~$10^{18}-10^{19}$   & $10^{-9}-10^{-8}$    \\
		\hline
	\end{tabular}
	\caption{The frequency range of the different gravitational wave detectors and the relevant mass range of the gravitational wave corresponding to the massive spin-2 KK mode excitations with mass $\text{Re}(m_{1})$.\label{tab4}}
\end{table}

In addition to the frequency, the lifetime of the massive KK graviton on the brane, obtained from the imaginary part of the QNM frequency, is crucial for its detectability. Only long-lived KK gravitons can travel far enough to be detected by gravitational wave detectors. Table~\ref{tab5} presents the prospect of KK graviton detection in various brane world scenarios. It can be seen that, for the RS-II thin brane and thick branes with small $s$ and $\delta$, the first QNM frequencies are extremely high, and the lifetimes are very short, making their detection challenging with low- and medium-frequency gravitational wave detectors. However, for thick branes with a large degree of splitting, the first QNM frequency may fall within the detection range of high-frequency gravitational wave detectors such as CAST. Therefore, compared with thin brane or single thick brane, thick branes with sub-brane have richer properties and better observation prospects, we hope to search for these extra-dimensional messengers in high-frequency gravitational wave detectors.

\begin{table}[htbp]
	\begin{tabular}{|c|c|c|c|}
		\hline
		&
		$m_{1}/k$ &
		$\omega_{1}$(Hz) &
		lifetime (s)	\\
		\hline
		RS-II brane  &~ $-0.419-0.577i$  & $10^{22}$ &~ $10^{-13}$    \\
		\hline
		$\delta=s=1$  &~$0.99702-0.52636i$   &  $10^{23}$ &~ $10^{-13}$   \\
		\hline
		$\delta=3,s=5$  &~$0.45005-0.01662i$   & $10^{22}$ &~ $10^{-11}$    \\
		\hline
		$\delta=5,s=35$  &~$0.28564-0.00180i$   & $10^{22}$ &~ $10^{-10}$    \\
		\hline
		$\delta=10,s=35$  &~$0.15048-0.00013i$   & $10^{22}$ &~ $10^{-9}$    \\
		\hline
		$\delta=20,s=35$  &~$0.07734-0.000012i$   & $10^{21}$ &~ $10^{-8}$    \\
		\hline
	\end{tabular}
	\caption{The frequency and lifetime of the first QNM for branes with different models and parameters. The extra dimension scale $k$ is set to $k=10^{-3}$eV.\label{tab5}}
\end{table}

\section{Conclusion and discussion}
\label{Conclusion}
In this paper, we investigated the gravitational QNMs and echoes of the split thick brane. The splitting of the brane results in a richer phenomenology. The quasinormal spectrum of the split thick brane is very different from that of the unsplit ones, especially the high overtone mode. As the degree of splitting increases, echo signals and a beating effect appear, which are unique features of thick branes with sub-brane structures.

We first review the split thick brane model and the corresponding transverse-traceless tensor fluctuations. We then derive the wave equation and the Schrodinger-like equation along the extra dimension for spin-2 KK gravitons on the brane. Based on the above equations, we study the quasinormal spectrum of the split thick brane in the frequency domain by the Bernstein spectral method, asymptotic iteration method, and direct integration method. The results are shown in Figs.~\ref{figms}, \ref{qnmplotten}, Tabs.~\ref{tab1}, ~\ref{tab2}, and ~\ref{tab3}. We find that the spectrum for $s > 1$ (corresponding to a split brane) is completely different from that for $s=1$ (corresponding to an unsplit brane). However, when $s$ continues to increase, the fundamental mode changes very little, but the high overtone modes change greatly. This indicates that the fundamental mode is relatively stable to the change of potential, while the high overtone modes are extremely sensitive to the change of potential, suggesting that the high overtone modes may be the key to identify the thick brane with little difference in configuration. To complement the results in the frequency domain, we also perform time-domain evolution. The evolution results are shown in Figs.~\ref{qnmsploteven} and~\ref{qnmsplotodd1}. The frequency of the first QNM obtained by numerical evolution is consistent with the results in the frequency domain. Finally, we study the gravitational echo of the thick brane with $\delta>1$. The results show that echo signals appear for larger values of $\delta$. This is because with the increase of $\delta$, the splitting degree of the thick brane also increases, the effective potential is similar to a resonant cavity, and the gravitational KK mode oscillates and leaks in the quasi-well, forming an echo. In addition, we also found that the different KK modes of the thick brane interfered with each other, resulting in the beating effect. This lays a foundation for further research on the co-evolution of multiple KK modes on the brane. In general, the splitting of the brane enriches the structure of extra dimensions and also leads to more interesting phenomenology. We hope to detect these extra-dimensional signals, especially echo signals, in future gravitational wave detection experiments.

Our work could be enhanced by several ways. For example, considering thick brane models with more sub-branes or brane arrays. The characteristic model of thin brane model with sub-brane structure is also worth studying. Characteristic modes and echoes for other test fields should also be investigated.

\section*{Acknowledgements}
This work was supported by the National Natural Science Foundation of China (Grants No. 12035005, No. 12405055, and No. 12347111), the China Postdoctoral Science Foundation (Grant No. 2023M741148), the Postdoctoral Fellowship Program of CPSF (Grant No. GZC20240458), and the National Key Research and Development Program of China (Grant No. 2020YFC2201400).


\begin{thebibliography}{99}
	
	
	
	
\bibitem{ArkaniHamed:1998rs}
N.~Arkani-Hamed, S.~Dimopoulos, and G.~R. Dvali, \emph{{The Hierarchy problem
		and new dimensions at a millimeter}},
{\emph{Phys. Lett. B}
	{\bfseries 429}, 263 (1998)},
[{{\ttfamily arXiv:hep-ph/9803315}}].


\bibitem{Antoniadis:1998ig}
I.~Antoniadis, N.~Arkani-Hamed, S.~Dimopoulos, and G.~R.~Dvali,
\emph{{New dimensions at a millimeter to a Fermi and superstrings at a TeV,}}
{\emph{ Phys. Lett. B}
	{\bfseries 436}, 257 (1998)},
[{{\ttfamily arXiv:hep-ph/9804398}}].



\bibitem{Randall:1999ee}
L.~Randall and R.~Sundrum, \emph{{A Large mass hierarchy from a small extra dimension}},
{\emph{Phys. Rev. Lett.}
	{\bfseries 83},  3370 (1999)},
[{{\ttfamily arXiv:hep-ph/9905221}}].



\bibitem{Randall:1999vf}
L.~Randall and R.~Sundrum, \emph{{An Alternative to compactification}},
{\emph{Phys. Rev. Lett.}
	{\bfseries 83},  4690 (1999)},
[{{\ttfamily arXiv:hep-th/9906064}}].


\bibitem{Shiromizu:1999wj}
T.~Shiromizu, K.~Maeda, and M.~Sasaki,
{\emph{The Einstein equation on the 3-brane world}},
{\emph{Phys. Rev. D} {\bfseries 62},  024012 (2000)},
[{{\ttfamily arXiv:gr-qc/9910076}}].


\bibitem{Tanaka:2002rb}
T.~Tanaka,
{\emph{Classical black hole evaporation in Randall-Sundrum infinite brane world}},
{\emph{Prog. Theor. Phys. Suppl.} {\bfseries 148},  307 (2003)},
[{{\ttfamily arXiv:gr-qc/0203082}}].

\bibitem{Gregory:2008rf}
R.~Gregory,
{\emph{Braneworld black holes}},
{\emph{Lect. Notes Phys.}  {\bfseries 769},  259 (2009)},
[{{\ttfamily arXiv:0804.2595}}].


\bibitem{Jaman:2018ucm}
N.~Jaman and K.~Myrzakulov,
{\emph{Braneworld inflation with an effective $\alpha$-attractor potential}},
{\emph{Phys. Rev. D} {\bfseries 99}, 103523 (2019)},
[{{\ttfamily arXiv:1807.07443}}].


\bibitem{Adhikari:2020xcg}
R.~Adhikari, M.~R.~Gangopadhyay, and Yogesh,
{\emph{Power Law Plateau Inflation Potential In The RS $II$ Braneworld Evading Swampland Conjecture}},
{\emph{Eur. Phys. J. C} {\bfseries 80}, 899 (2020)},
[{{\ttfamily arXiv:2002.07061}}].



\bibitem{Geng:2020fxl}
H.~Geng, A.~Karch, C.~Perez-Pardavila, S.~Raju, L.~Randall, M.~Riojas, and S.~Shashi,
{\emph{Information Transfer with a Gravitating Bath}},
{\emph{SciPost Phys.} {\bfseries 10} 103 (2021)} , 
[{{\ttfamily arXiv:2012.04671}}].



\bibitem{Geng:2021iyq}
H.~Geng, S.~L\"ust, R.~K.~Mishra, and D.~Wakeham,
{\emph{Holographic BCFTs and Communicating Black Holes}},
{\emph{JHEP} {\bfseries 08} 003 (2021)}, 
[{{\ttfamily arXiv:2104.07039}}].


\bibitem{Geng:2022dua}
H.~Geng, L.~Randall, and E.~Swanson,
{\emph{BCFT in a black hole background: an analytical holographic model}},
{\emph{JHEP} {\bfseries 12}, 056  (2022)},
[{{\ttfamily arXiv:2209.02074}}].


\bibitem{Bhattacharya:2021jrn}
A.~Bhattacharya, A.~Bhattacharyya, P.~Nandy, and A.~K.~Patra,
{\emph{Islands and complexity of eternal black hole and radiation subsystems for a doubly holographic model}},
{\emph{JHEP} {\bfseries 05}, 135 (2021)},
[{{\ttfamily arXiv:2103.15852}}].
	
	

\bibitem{ValeixoBento:2022qca}
B.~Valeixo Bento, D.~Chakraborty, S.~Parameswaran, and I.~Zavala,
{\emph{Gravity at the tip of the throat}},
{\emph{JHEP} {\bfseries 09}, 208  (2022)},
[{{\ttfamily arXiv:2204.02086}}].

	
	

\bibitem{DeWolfe:1999cp}
O.~DeWolfe, D.~Z.~Freedman, S.~S.~Gubser, and A.~Karch,
\emph{{Modeling the fifth-dimension with scalars and gravity}},
{\emph{Phys. Rev. D} {\bfseries 62}, 046008 (2000)},
[{{\ttfamily arXiv:hep-th/9909134}}].


\bibitem{Gremm:1999pj}
M.~Gremm,
\emph{{Four-dimensional gravity on a thick domain wall}},
{\emph{Phys. Lett. B} {\bfseries 478}, 434 (2000)},
[{{\ttfamily arXiv:hep-th/9912060}}].


\bibitem{Csaki:2000fc}
C.~Csaki, J.~Erlich, T.~J.~Hollowood, and Y.~Shirman,
\emph{{Universal aspects of gravity localized on thick branes}},
{\emph{Nucl. Phys. B} {\bfseries 581}, 309 (2000)},
[{{\ttfamily arXiv:hep-th/0001033}}].



\bibitem{Akama:1982jy}
K.~Akama,
\emph{{An Early Proposal of `Brane World'}},
{\emph{Lect. Notes Phys.} {\bfseries 176}, 267 (1982)},
[{{\ttfamily arXiv:hep-th/0001113}}].


\bibitem{Rubakov:1983bb}
V.~A.~Rubakov and M.~E.~Shaposhnikov,
\emph{{Do We Live Inside a Domain Wall?}}
{\emph{Phys. Lett. B} {\bfseries 125}, 136 (1983)}.


\bibitem{Afonso:2007gc}
V.~I.~Afonso, D.~Bazeia, R.~Menezes, and A.~Y.~Petrov,
\emph{{f(R)-Brane}},
{\emph{Phys. Lett. B} {\bfseries 658}, 71 (2007)},
[{{\ttfamily arXiv:0710.3790}}].




\bibitem{Dzhunushaliev:2010fqo}
V.~Dzhunushaliev and V.~Folomeev,
\emph{{Spinor brane}},
{\emph{Gen. Rel. Grav.} {\bfseries 43}, 1253  (2011)},
[{{\ttfamily arXiv:0909.2741}}].



\bibitem{Dzhunushaliev:2011mm}
V.~Dzhunushaliev and V.~Folomeev,
\emph{{Thick brane solutions supported by two spinor fields}},
{\emph{Gen. Rel. Grav.} {\bfseries 44}, 253  (2012)},
[{{\ttfamily arXiv:1104.2733}}].


\bibitem{Geng:2015kvs}
W.-J.~Geng and H.~Lu,
\emph{{Einstein-Vector Gravity, Emerging Gauge Symmetry and de Sitter Bounce}},
{\emph{Phys. Rev. D} {\bfseries 93}, 044035  (2016)},
[{{\ttfamily arXiv:1511.03681}}].



\bibitem{Melfo2006}
A.~Melfo, N.~Pantoja, and J.~D. Tempo, {\it {Fermion localization on thick branes}},  {\em Phys. Rev. D} {\bf 73},  044033 (2006), [{{\tt arXiv:hep-th/0601161}}].

\bibitem{Almeida2009}
C.~A. Almeida, R.~Casana, M.~M. Ferreira, and A.~R. Gomes, {\it {Fermion localization and resonances on two-field thick branes}},  {\em Phys. Rev. D} {\bf 79},  125022 (2009), [{{\tt arXiv:0901.3543}}].

\bibitem{Zhao2010}
Z.-H. Zhao, Y.-X. Liu, and H.-T. Li, {\it {Fermion localization on asymmetric two-field thick branes}},  {\em Class. Quantum Gravity} {\bf 27},  185001 (2010), [{{\tt arXiv:0911.2572}}].

\bibitem{Chumbes2011}
A.~E.~R. Chumbes, A.~E.~O. Vasquez, and M.~B. Hott, {\it {Fermion localization on a split brane}},  {\em Phys. Rev. D} {\bf 83},  105010 (2011), [{{\tt arXiv:1012.1480}}].

\bibitem{Liu2011}
Y.-X. Liu, Y.~Zhong, Z.-H. Zhao, and H.-T. Li,
{\it {Domain wall brane in squared curvature gravity}},  
{\em J. High Energy Phys.} {\bf 2011}, 135 (2011), 
[{{\tt arXiv:1104.3188v2}}].



\bibitem{Bazeia:2013uva}
D.~Bazeia, A.~S.~Lob\~ao, Jr., R.~Menezes, A.~Y.~Petrov, and A.~J.~da Silva,
{\it {Braneworld solutions for F(R) models with non-constant curvature}},
{\em Phys. Lett. B} {\bf 729}, 127 (2014),
[{{\tt arXiv:1311.6294}}].



\bibitem{Xie2017}
Q.-Y. Xie, H.~Guo, Z.-H. Zhao, Y.-Z. Du, and Y.-P. Zhang, {\it {Spectrum structure of a fermion on Bloch branes with two scalar-fermion couplings}}, {\em Class. Quantum Gravity} {\bf 34},  055007 (2017), [{{\tt arXiv:1510.03345}}].

\bibitem{Gu2017}
B.-M. Gu, Y.-P. Zhang, H.~Yu, and Y.-X. Liu, {\it {Full linear perturbations and localization of gravity on $f(R, T)$ brane}},  {\em Eur. Phys. J. C} {\bf 77}, 115 (2017), [{{\tt arXiv:1606.07169}}].

\bibitem{ZhongYuan2017}
Y.~Zhong and Y.-X. Liu, {\it {Linearization of a warped $f(R)$ theory in the higher-order frame}},  {\em Phys. Rev. D} {\bf 95}, 104060 (2017),
[{{\tt arXiv:1611.08237}}].

\bibitem{ZhongYuan2017b}
Y.~Zhong, K.~Yang, and Y.-X. Liu, {\it {Linearization of a warped $f(R)$ theory in the higher-order frame II: The equation of motion approach}}, {\em Phys. Rev. D} {\bf 97},  044032 (2017), [{{\tt arXiv:1708.03737}}].

\bibitem{Zhou2018}
X.-N. Zhou, Y.-Z. Du, H.~Yu, and Y.-X. Liu,
{\it {Localization of gravitino field on $f(R)$-thick branes}},
{\em Sci. China Physics, Mech. Astron.} {\bf 61}, 110411 (2018),
[{{\tt arXiv:1703.10805}}].



\bibitem{Hendi:2020qkk}
S.~H.~Hendi, N.~Riazi, and S.~N.~Sajadi,
{\it {$Z_2$-symmetric thick brane with a specific warp function}},
{\em Phys. Rev. D} {\bf 102}, 124034 (2020),
[{{\tt arXiv:2011.11093}}].



\bibitem{Xie:2021ayr}
Q.-Y.~Xie, Q.-M.~Fu, T.-T.~Sui, L.~Zhao, and Y.~Zhong,
{\it {First-Order Formalism and Thick Branes in Mimetic Gravity}},
{\em Symmetry} {\bf 13}, 1345 (2021),
[{{\tt arXiv:2102.10251}}].



\bibitem{Moreira:2021uod}
A.~R.~P.~Moreira, F.~C.~E.~Lima, J.~E.~G.~Silva, and C.~A.~S.~Almeida,
{\it {First-order formalism for thick branes in $f(T,{\mathscr {T}})$ gravity}},
{\em Eur. Phys. J. C}   {\bf 81}, 1081 (2021),
[{{\tt arXiv:2107.04142}}].


\bibitem{Xu:2022ori}
N.~Xu, J.~Chen, Y.-P.~Zhang, and Y.-X.~Liu,
{\it {Multi-kink brane in Gauss-Bonnet gravity}},
[{{\tt arXiv:2201.10282}}].


\bibitem{Silva:2022pfd}
J.~E.~G.~Silva, R.~V.~Maluf, G.~J.~Olmo, and C.~A.~S.~Almeida,
{\it {Braneworlds in $f(Q)$ gravity}},
[{{\tt arXiv:2203.05720}}].


\bibitem{Xu:2022gth}
Y.-Q.~Xu and X.-D.~Zhang,
{\it {Tensor Perturbations and Thick Branes in Higher Dimensional Gauss-Bonnet Gravity}},
[{{\tt arXiv:2203.13401}}].


\bibitem{Dzhunushaliev:2009va}
V.~Dzhunushaliev, V.~Folomeev, and M.~Minamitsuji,
\emph{{Thick brane solutions}},
{\emph{Rept. Prog. Phys.}} {\bfseries 73}, 066901  (2010),
[{{\tt arXiv:0904.1775}}].

\bibitem{Maartens:2010ar}
R.~Maartens and K.~Koyama,
{\it {Brane-World Gravity}},
{\em Living Rev. Rel.} {\bf 13}, 5 (2010),
[{{\tt arXiv:1004.3962}}].

\bibitem{Liu:2017gcn}
Y.-X.~Liu,
{\it {Introduction to Extra Dimensions and Thick Braneworlds}},
[{{\tt arXiv:1707.08541}}].


\bibitem{Ahluwalia:2022ttu}
D.~V.~Ahluwalia, J.~M.~H.~da Silva, C.~Y.~Lee, Y.-X.~Liu, S.~H.~Pereira, and M.~M.~Sorkhi,
\emph{{Mass dimension one fermions: Constructing darkness}},
{\emph{Phys. Rept.}} {\bfseries 967}, 1 (2022),
[{{\tt arXiv:2205.04754}}].








\bibitem{Berti:2009kk}
E.~Berti, V.~Cardoso, and A.~O.~Starinets,
\emph{{Quasinormal modes of black holes and black branes}},
{\emph{Class. Quant. Grav.} {\bfseries 26},
	163001 (2009)},
[{{\ttfamily arXiv:0905.2975}}].


\bibitem{Kokkotas:1999bd}
K.~D.~Kokkotas and B.~G.~Schmidt,
\emph{{Quasinormal modes of stars and black holes}},
{\emph{Living Rev. Rel.} {\bfseries 2}, 2 (1999)},
[{{\ttfamily  arXiv:gr-qc/9909058}}].


\bibitem{Nollert:1999ji}
H.~P.~Nollert,
\emph{{TOPICAL REVIEW: Quasinormal modes: the characteristic `sound' of black holes and neutron stars}},
{\emph{Class. Quant. Grav.} {\bfseries 16}, R159 (1999)}.


\bibitem{Konoplya:2011qq}
R.~A.~Konoplya and A.~Zhidenko,
\emph{{Quasinormal modes of black holes: From astrophysics to string theory}},
{\emph{ Rev. Mod. Phys.} {\bfseries 83}, 793 (2011)},
[{{\ttfamily  arXiv:1102.4014}}].


\bibitem{Cardoso:2016rao}
V.~Cardoso, E.~Franzin, and P.~Pani,
\emph{{Is the gravitational-wave ringdown a probe of the event horizon?}}
{\emph{ Phys. Rev. Lett.} {\bfseries 116}, 171101  (2016)},
[erratum: \emph{ Phys. Rev. Lett.}  {\bfseries 117} , 089902 (2016)]
[{{\ttfamily  arXiv:1602.07309}}].



\bibitem{Jusufi:2020odz}
K.~Jusufi, M.~Azreg-A\"\i{}nou, M.~Jamil, S.-W.~Wei, Q.~Wu, and A.-Z.~Wang,
\emph{{Quasinormal modes, quasiperiodic oscillations, and the shadow of rotating regular black holes in nonminimally coupled Einstein-Yang-Mills theory}},
{\emph{ Phys. Rev. D}  {\bfseries 103},  024013 (2021)},
[{{\ttfamily  arXiv:2008.08450}}].


\bibitem{Cheung:2021bol}
M.~H.~Y.~Cheung, K.~Destounis, R.~P.~Macedo, E.~Berti, and V.~Cardoso,
\emph{{Destabilizing the Fundamental Mode of Black Holes: The Elephant and the Flea}},
{\emph{ Phys. Rev. Lett.}  {\bfseries 128}, 111103  (2022)},
[{{\ttfamily arXiv:2111.05415}}].


\bibitem{Seahra:2005wk}
S.~S.~Seahra,
\emph{Ringing the Randall-Sundrum braneworld: Metastable gravity wave bound states},
{\emph{Phys. Rev. D} {\bfseries  72}, 066002 (2005)},
[{{\ttfamily arXiv:hep-th/0501175}}].


\bibitem{Seahra:2005iq}
S.~S.~Seahra,
\emph{Metastable massive gravitons from an infinite extra dimension},
{\emph{Int. J. Mod. Phys. D} {\bfseries  14},  2279 (2005)},
[{{\ttfamily arXiv:hep-th/0505196}}].


\bibitem{Tan:2022vfe}
Q.~Tan, W.-D.~Guo, and Y.-X.~Liu,
\emph{Sound from extra dimension: quasinormal modes of thick brane},
{\emph{Phys. Rev. D } {\bfseries  106}, 044038 (2022)},
[{{\ttfamily arXiv:2205.05255}}].





\bibitem{Tan:2023cra}
Q.~Tan, W.-D.~Guo, Y.-P.~Zhang, and Y.-X.~Liu,
{\it {Characteristic modes of a thick brane: Resonances and quasinormal modes}},
{\em Phys. Rev. D} {\bf 109}, 024017 (2024),
[{{\tt arXiv:2304.09363}}].




\bibitem{Tan:2024url}
Q.~Tan, Y.~Zhong, and W.-D.~Guo,
{\it {Quasibound and quasinormal modes of a thick brane in Rastall gravity}},
{\em JHEP} {\bf 07}, 252 (2024),
[{{\tt arXiv:2404.11217}}].

\bibitem{Jia:2024pdk}
H.-L.~Jia, W.-D.~Guo, Q.~Tan, and Y.-X.~Liu,
{\it {Quasinormal ringing of thick braneworlds with a finite extra dimension}},
{\em Phys. Rev. D} {\bf 110}, 064077 (2024),
[{{\tt arXiv:2406.03929}}].

\bibitem{Tan:2024aym}
Q.~Tan, S.~Long, W.-K.~Deng, and J.-J.~Jing,
{\it {Graviscalar quasinormal modes and asymptotic tails of a thick brane}},
[{{\tt arXiv:2409.06947}}].

\bibitem{Melfo:2002wd}
A.~Melfo, N.~Pantoja, and A.~Skirzewski,
{\it {Thick domain wall space-times with and without reflection symmetry}},
{\em Phys. Rev. D} {\bf 67}, 105003 (2003),
[{{\tt arXiv:gr-qc/0211081}}].

\bibitem{Castillo-Felisola:2004omi}
O.~Castillo-Felisola, A.~Melfo, N.~Pantoja, and A.~Ramirez,
{\it {Localizing gravity on exotic thick three-branes}},
{\em Phys. Rev. D} {\bf 70}, 104029 (2004),
[{{\tt arXiv:hep-th/0404083}}].





\bibitem{deBrito:2014pqa}
G.~P.~de Brito, R.~A.~C.~Correa, and A.~de Souza Dutra,
{\it {Analytical multikinks in smooth potentials}},
{\em Phys. Rev. D} {\bf 89}, 065039 (2014),
[{{\tt arXiv:1403.2951}}].



\bibitem{deSouzaDutra:2014ddw}
A.~de Souza Dutra, G.~P.~de Brito, and J.~M.~Hoff da Silva,
{\it {Method for obtaining thick brane models}},
{\em Phys. Rev. D} {\bf 91}, 086016 (2015),
[{{\tt arXiv:1412.5543}}].

\bibitem{Farokhtabar:2016fhm}
A.~Farokhtabar and A.~Tofighi,
{\it {Localization of Massive and Massless Fermions on Two-Field Branes}},
{\em Adv. High Energy Phys.} {\bf 2017}, 3926286 (2017),
[{{\tt arXiv:1610.00899}}].

\bibitem{Xie:2019jkq}
Q.-Y.~Xie, Z.-H.~Zhao, J.~Yang, and K.~Yang,
{\it {Fermion Localization and Degenerate Resonances on Brane Array}},
{\em Class. Quant. Grav.} {\bf 37}, 025012 (2020),
[{{\tt arXiv:1901.11253}}].



\bibitem{Guerrero:2006gj}
R.~Guerrero, A.~Melfo, N.~Pantoja, and R.~O.~Rodriguez,
{\it {Close to the edge: Hierarchy in a double braneworld}},
{\em Phys. Rev. D} {\bf 74}, 084025 (2006),
[{{\tt arXiv:hep-th/0605160}}].

\bibitem{Ahmed:2012nh}
A.~Ahmed and B.~Grzadkowski,
{\it {Brane modeling in warped extra-dimension}},
{\em JHEP} {\bf 01}, 177 (2013),
[{{\tt arXiv:1210.6708}}].

\bibitem{deSouzaDutra:2013rwa}
A.~de Souza Dutra, G.~P.~de Brito, and J.~M.~Hoff da Silva,
{\it {Asymmetrical bloch branes and the hierarchy problem}},
{\em EPL} {\bf 108}, 11001 (2014),
[{{\tt arXiv:1312.0091}}].










\bibitem{Cardoso:2017cqb}
V.~Cardoso and P.~Pani,
{\it {Tests for the existence of black holes through gravitational wave echoes}},
{\em Nature Astron.} {\bf 1}, 586 (2017),
[{{\tt  arXiv:1709.01525}}].




\bibitem{Cardoso:2019rvt}
V.~Cardoso and P.~Pani,
{\it {Testing the nature of dark compact objects: a status report}},
{\em Living Rev. Rel.} {\bf 22}, 4 (2019),
[{{\tt arXiv:1904.05363}}].

\bibitem{Mark:2017dnq}
Z.~Mark, A.~Zimmerman, S.-M.~Du, and Y.~Chen,
{\it {A recipe for echoes from exotic compact objects}},
{\em Phys. Rev. D} {\bf 96}, 084002 (2017),
[{{\tt arXiv:1706.06155}}].

\bibitem{Conklin:2017lwb}
R.~S.~Conklin, B.~Holdom, and J.~Ren,
{\it {Gravitational wave echoes through new windows}},
{\em Phys. Rev. D} {\bf 98}, 044021 (2018),
[{{\tt arXiv:1712.06517}}].

\bibitem{Barcelo:2017lnx}
C.~Barcel'o, R.~Carballo-Rubio, and L.~J.~Garay,
{\it {Gravitational wave echoes from macroscopic quantum gravity effects}},
{\em JHEP} {\bf 05}, 054 (2017),
[{{\tt arXiv:1701.09156}}].

\bibitem{Qian:2024zvq}
W.-L.~Qian, Q.-Y.~Pan, B.~Wang, and R.-H.~Yue,
{\it {Late-time tail and echoes of Damour-Solodukhin wormholes}},
{\em Phys. Lett. B} {\bf 856}, 138874 (2024),
[{{\tt arXiv:2402.05485}}].

\bibitem{Lin:2023qgd}
K.~Lin,
{\it {Quasinormal modes and echo effect of a cylindrical anti--de Sitter black hole spacetime with a thin shell}},
{\em Phys. Rev. D} {\bf 107}, 124002 (2023),
[{{\tt arXiv:2306.01269}}].










\bibitem{Zhu:2024gvl}
C.-C.~Zhu, J.~Chen, W.-D.~Guo and Y.-X.~Liu,
{\it {Gravitational Echoes from Braneworlds}},
{{\tt arXiv:2406.16256}}.



\bibitem{Cooper:1994eh}
F.~Cooper, A.~Khare, and U.~Sukhatme,
\emph{{Supersymmetry and quantum mechanics}},
{\emph{Phys. Rept. }{\bfseries 251},  267 (1995)},
[{{\ttfamily arXiv:hep-th/9405029}}].



\bibitem{Ge:2018vjq}
B.-X.~Ge, J.~Jiang, B.~Wang, H.-B.~Zhang, and Z.~Zhong,
\emph{Strong cosmic censorship for the massless Dirac field in the Reissner-Nordstrom-de Sitter spacetime},
{\emph{JHEP} {\bfseries  01}, 123 (2019)},
[{{\ttfamily arXiv:1810.12128}}].





\bibitem{Fortuna:2020obg}
S.~Fortuna and I.~Vega,
\emph{Bernstein spectral method for quasinormal modes and other eigenvalue problems},
{\emph{Eur. Phys. J. C} {\bfseries  83}, 1170 (2023)},
[{{\ttfamily arXiv:2003.06232}}].


\bibitem{Pani:2013pma}
P.~Pani,
\emph{Advanced Methods in Black-Hole Perturbation Theory},
{\emph{Int. J. Mod. Phys. A} {\bfseries  28}, 1340018 (2013)},
[{{\ttfamily arXiv:1305.6759}}].


\bibitem{Cho:2011sf}
H.-T.~Cho, A.~S.~Cornell, J.~Doukas, T.-R.~Huang, and W.~Naylor,
\emph{A New Approach to Black Hole Quasinormal Modes: A Review of the Asymptotic Iteration Method},
{\emph{Adv. Math. Phys.} {\bfseries  2012}, 281705 (2012)},
[{{\ttfamily arXiv:1111.5024}}].



\bibitem{Grandclement:2007sb}
P.~Grandclement and J.~Novak,
{\it {Spectral methods for numerical relativity}},''
{\em Living Rev. Rel.}  {\bf 12}, 1 (2009),
[{{\tt arXiv:0706.2286}}].

\bibitem{Jansen:2017oag}
A.~Jansen,
{\it {Overdamped modes in Schwarzschild-de Sitter and a Mathematica package for the numerical computation of quasinormal modes}},
{\em Eur. Phys. J. Plus}  {\bf 132}, 546   (2017),
[{{\tt arXiv:1709.09178}}].


\bibitem{Jaramillo:2020tuu}
J.~L.~Jaramillo, R.~Panosso Macedo, and L.~Al Sheikh,
{\it {Pseudospectrum and Black Hole Quasinormal Mode Instability}},
{\em Phys. Rev. X} {\bf 11}, 031003 (2021),
[{{\tt arXiv:2004.06434}}].




\bibitem{Chung:2023wkd}
A.~K.~W.~Chung, P.~Wagle, and N.~Yunes,
{\it {Spectral method for metric perturbations of black holes: Kerr background case in general relativity}},
{\em Phys. Rev. D}  {\bf109}, 044072  (2024),
[{{\tt  arXiv:2312.08435}}].


\bibitem{Witek:2012tr}
H.~Witek, V.~Cardoso, A.~Ishibashi, and U.~Sperhake,
{\it {Superradiant instabilities in astrophysical systems}},
{\em Phys. Rev. D} {\bf 87}, 043513 (2013),
[{{\tt arXiv:1212.0551}}].





\bibitem{Lisa:2017}
Amaro-Seoane P, Audley H, and Babak S, \textit{et al.}
{\it {Laser interferometer space antenna}},
arXiv:1702.00786.



\bibitem{LIGOScientific:2016aoc}
B.~P.~Abbott \textit{et al.} [LIGO Scientific and Virgo],
{\it {Observation of Gravitational Waves from a Binary Black Hole Merger}},
{\em Phys. Rev. Lett. }  {\bf 116}, 061102 (2016),
[{{\tt arXiv:1602.03837}}].



\bibitem{Ejlli:2019bqj}
A.~Ejlli, D.~Ejlli, A.~M.~Cruise, G.~Pisano, and H.~Grote,
{\it {Upper limits on the amplitude of ultra-high-frequency gravitational waves from graviton to photon conversion}},
{\em Eur. Phys. J. C}  {\bf 79}, 1032  (2019),
[{{\tt arXiv:1908.00232}}].




\bibitem{Aggarwal:2020olq}
N.~Aggarwal, O.~D.~Aguiar, A.~Bauswein, G.~Cella, S.~Clesse, A.~M.~Cruise, V.~Domcke, D.~G.~Figueroa, A.~Geraci, and M.~Goryachev, \textit{et al.}
{\it {Challenges and opportunities of gravitational-wave searches at MHz to GHz frequencies}},
{\em Living Rev. Rel.}  {\bf 24}, 4 (2021),
[{{\tt arXiv:2011.12414}}].




\end{thebibliography}
\end{document}